\documentclass{aa}  

\usepackage[T1]{fontenc}
\usepackage{graphicx}
\usepackage{txfonts}
\usepackage{comment}
\usepackage{xcolor}
\usepackage[percent]{overpic}
\usepackage{academicons}
\newcommand{\orcid}[1]{}

\usepackage[colorlinks=true,linkcolor=bluea,allcolors=blue]{hyperref}

\begin{document} 
\newcommand{\ngc}[1]{NGC\,#1} 
\newcommand{\water}{H$_2$O} 
\newcommand{\rottrans}[3]{#1~$(J=#2\rightarrow#3)$} 
\newcommand{\kms}{km\,s$^{-1}$}
\newcommand{\mone}{$^{-1}$}
\newcommand{\mJybeam}{mJy\,beam$^{-1}$}

\newcommand{\PvdW}[1]{\textcolor{red}{\textbf{[PvdW: #1]}}}
\newcommand{\VGR}[1]{\textcolor{blue}{\textbf{[VGR: #1]}}}
\newcommand{\jack}[1]{\textcolor{orange}{\textbf{[Jack: #1]}}}
\newcommand{\PvdWb}[1]{\textcolor{purple}{\textbf{[PvdW2: #1]}}}

   \title{Decoding the molecular torus of NGC 1068: }
   \subtitle{Insights into its structure and kinematics from high-resolution ALMA observations}
   \author{V. G\'amez Rosas\inst{1}\orcid{0000-0002-8830-6199}
          \and
          P. van der Werf\inst{1}\orcid{0000-0001-5434-5942}
         \and
        J. F. Gallimore\inst{2}\orcid{0000-0002-6972-2760}
        \and
        V. Impellizzeri\inst{1}\orcid{0000-0002-3443-2472}
        \and
        W. Jaffe\inst{1}
        \and 
        S. García-Burillo\inst{4}\orcid{0000-0003-0444-6897}
        \and 
        S. Aalto\inst{8}
        \and 
        L. Burtscher\inst{1}\orcid{0000-0003-1014-043X}
        \and 
        V. Casasola\inst{5}\orcid{0000-0002-3879-6038}
        \and 
        F. Combes\inst{7}\orcid{0000-0003-2658-7893}
        \and 
        C. Henkel\inst{10}
        \and 
        I. Márquez\inst{6}\orcid{0000-0003-2629-1945}
        \and 
        S. Martín\inst{9}$^{,}$\inst{12}\orcid{0000-0001-9281-2919}
        \and 
        C. Ramos Almeida\inst{3}$^{,}$\inst{11}\orcid{0000-0001-8353-649X}
        \and 
        S. Viti\inst{1}\orcid{0000-0001-8504-8844}
        }
   \institute{Leiden Observatory,
              Leiden University, P.O. Box 9513, 2300 RA Leiden, The Netherlands
         \and
              Department of Physics and Astronomy, Bucknell University, One Dent Drive Lewisburg, PA 17837, USA
        \and
             Instituto de Astrofísica de Canarias  Calle Vía Láctea, s/n E-38205 La Laguna, Tenerife, Spain
         \and
            Observatorio Astronómico Nacional Aptdo 1143 28800 Alcalá de Henares, Spain
        \and
            INAF Istituto di Radioastronomia Via Piero Gobetti 101 40129 Bologna, Italy
        \and
             Inst de Astrofisica de Andalucía Extragalactic Astronomy Glorieta de la astronomia, S/N 18008 Granada, Spain
        \and
             Observatoire de Paris, LERMA, Collège de France, CNRS, PSL University, Sorbonne University, 75014, Paris, France
        \and             
             Chalmers University of Technology Department of Earth and Space Sciences Onsala Space Observatory 439 92 Onsala, Sweden
         \and            
             European Southern Observatory, Alonso de C{\'o}rdova, 3107, Vitacura, Santiago 763-0355, Chile
         \and 
             MPI für Radioastronomie Auf dem H{\"u}gel 69, 53121 Bonn, Germany
         \and              
             Departamento de Astrof\' isica, Universidad de La Laguna, E-38206, La Laguna, Tenerife, Spain
        \and
            Joint ALMA Observatory, Alonso de C{\'o}rdova, 3107, Vitacura, Santiago 763-0355, Chile
             }

   \date{Accepted May 1, 2025}

  \abstract
{Previous kinematic analysis of the pc-scale molecular torus in NGC~1068 have revealed a very complex and inhomogeneous system involving several physical components including non-circular motions, turbulence, high-velocity outflows, and possibly even counter-rotation.}
{Our study aims to dissect the kinematics and morphology of the molecular gas within the near-nuclear region of NGC 1068 to understand the mechanisms in the central AGN that might be fuelling it, and the impact of its energy output on the surrounding molecular gas.}
{We present high angular and spectral resolution ALMA observations of the \rottrans{HCO$^+$}{4}{3} and \rottrans{CO}{3}{2} molecular lines in the near-nuclear region of the prototype Seyfert 2 galaxy NGC 1068. The spatial resolution (1.1~pc) is almost two times better than that of previous works studying the same molecular lines at the same transitions and is the highest resolution achievable with ALMA at these frequencies. Our analysis focuses on moment maps, position-velocity (PV) diagrams, and spectra obtained at the position of the nuclear continuum source, along with a simple kinematic model developed using the 3DBarolo software. }
{Our observations reveal significant asymmetry between the eastern and western sides of the nuclear disc in terms of morphology, velocity, and line intensity. 
The broad lines ($\sigma \sim$90 km/s) seen in the inner 2 pc could be accounted for by either beam smearing or highly turbulent gas in this region. 

Outside this radius the mean velocities drop to $\pm$30 km/s, which cannot be explained by asymmetric drift.  
We find low velocity connections extending to 13 pc suggesting interactions with larger scale structures. The CO/HCO$^+$ line ratio at the nucleus reported here are extremely low compared to values in the literature of the same galaxy at lower spatial resolutions.
We find high-velocity redshifted clouds in absorption and emission at the nuclear position.}
{The molecular environment near the nucleus of NGC 1068 is highly disturbed and asymmetric, marked by the presence of a high-velocity infalling cloud.
High excitation temperatures, high molecular column densities along with the unusually low CO/HCO$^+$ line ratio close to the nucleus seem to indicate intense interaction with AGN radiation. 
These findings underscore the complexity of AGN feeding mechanisms and the pivotal role of high-resolution studies in unravelling the physical processes at play near supermassive black holes.}
 

   \maketitle

\section{Introduction}
Spectropolarimetry reveals hidden broad-line regions in bright type 2 Seyfert nuclei \citep[e.g.][]{1985ApJ...297..621A,1990ApJ...355..456M,1992ApJ...397..452T,1995ApJ...440..565T,2000ApJ...540L..73M,2016MNRAS.461.1387R}. The central engine is therefore thought to be obscured by a ring of dusty molecular gas, the `obscuring torus,' located within the central few pc \citep{1988ApJ...329..702K}. In Seyfert unification schemes, all type 2 Seyfert galaxies are viewed from an unfavourable angle through the obscuring torus \citep{1993ARA&A..31..473A,1995PASP..107..803U}. However, to explain the fraction of type 2 Seyferts, the scale height of the torus needs to be comparable to the radius \citep{2005MNRAS.360..565S,2007ApJ...661...52K,2017ApJ...846..155K}. How the obscuring torus stably maintains such a scale height remains an unanswered question. The current thinking involves radiation pressure support or dusty outflows driven by radiation pressure \citep{1986ApJ...308L..55K,1995ApJ...450..628P,2008ApJ...675L...5D,2011ApJ...741...29D,2017ApJ...843...58C,2020ApJ...900..174V}, also referred to as radiation driven fountains (e.g., \cite{Wada2015}), nuclear starburst activity in the torus (e.g. \cite{WadaNorman2002}), or magnetohydrodynamic acceleration \citep{1992ApJ...385..460E,1994ApJ...434..446K,2006ApJ...648L.101E,2016ApJ...825...67C,2018A&A...615A.164V}. However, testing these ideas has been challenging because, at the distance of the nearest, bright Seyfert galaxies, $1~\mbox{pc} \la 15~\mbox{mas}$.

The galaxy \ngc{1068} is the ideal target for detailed studies of the obscuring torus. It is the closest luminous Seyfert~2 galaxy (D = 14.4 Mpc; 1\arcsec = 70 pc) and one of the first shown to have a hidden broad-line region \citep{1985ApJ...297..621A}. The X-ray spectrum shows Fe K$\alpha$ with a large equivalent width, indicating that the obscuring region is Compton-thick ($N_{H} \ga 10^{24}~\mbox{cm}^{-2}$; \citealt{2004A&A...414..155M}). Optical continuum polarimetry of the narrow-line region reveals a centro-symmetric pattern expected for Thomson scattering, but there is no bright source at the centre of symmetry \citep{1995ApJ...452L..87C,1999ApJ...518..676K}. The central engine of the AGN (Active Galactic Nucleus) produces a 100~pc scale jet seen in radio continuum, in which the compact, flat-spectrum radio source S1 marks the most likely position of the obscured AGN \citep{1996ApJ...464..198G,1996MNRAS.278..854M,1997Natur.388..852G,2004ApJ...613..794G}. There is a rotating ring of \water{} megamaser emission associated with S1 \citep{1996ApJ...462..740G,GG97,2001ApJ...556..694G}. The maser kinematics are consistent with a central mass $M \simeq 17\times 10^6~\mbox{M}_{\odot}$. 

High spatial resolution ($\sim 1-10$~mas) interferometric observations made with the Very Large Telescope Interferometer (VLTI) in the infrared have resolved the dusty regions close to the AGN with somewhat contradicting interpretations. 
On the one hand, the K band image reconstruction obtained with the data from the GRAVITY instrument finds a thin incomplete ring of hot ($\sim 1500~\mbox{K}$) dust with a radius of 0.24 pc oriented similar to the \water{} maser disc, which was associated to the dust sublimation radius \citep{2020A&A...634A...1G}.
On the other hand, image reconstruction obtained with the data from the instrument MATISSE along the L, M and N bands identified two dusty regions: (i) a two component asymmetric source of hot ($\sim 900~\mbox{K}$) dust interpreted as the upper and lower openings of a geometrically and optically thick torus extending to a maximum distance of $\sim1$pc from the AGN and (ii) a component of warm ($\sim 300~\mbox{K}$) dust extending over 3 -- 10 pc along the north-south axis which probably corresponds to the base of the polar dust emission in the ionisation cone \citep{2022Natur.602..403G}. 

Molecular lines afford the best probe of the kinematics of the obscuring region. Recently, the Atacama Large Millimeter Array (ALMA) has started to resolve the molecular gas associated with the central engine on pc-scales (cf. \cite{2016ApJ...822L..10I, 2018ApJ...859..144A, 2019A&A...628A..65A, 2018ApJ...853L..25I}). At the same time, the underlying continuum of the molecular lines provides information about the distribution of the cold dust or the areas of emission of thermal free-free radiation from ionised gas, or non-thermal synchroton radiation from  acceleration of ions in magnetic fields. Recent efforts are focussing on compiling a large sample of AGNs at high spatial resolution to better disentangle such mechanisms. \cite{GarciaBurillo2021} analyses ten nearby Seyfert galaxies selected from an ultra-hard X-ray survey with resolutions of approximately 0.1" (7-13 pc). The observations showed the presence of circumnuclear discs (CNDs) around the active galactic nuclei (AGN), with orientations that were predominantly perpendicular to the AGN wind axes, which is consistent with theoretical expectations for dusty molecular tori. \cite{Audibert2019} and \cite{Combes2019} reported the presence of molecular tori in six out of seven studied galaxies which do not align with the primary plane of the host galaxy's disc in several cases. Furthermore, in many cases, the AGN was found to be slightly off-centred with respect to the molecular torus. The variations in gas velocities and behaviours seen in the \rottrans {CO}{3}{2} line suggest that molecular tori undergo perturbations due to gravitational interactions \citep{Audibert2019}. This complex dynamic relationship suggests ongoing interactions among various components within the galaxy \citep{Audibert2019} and that the supermassive black hole can be wandering by several tens of parsecs around the centre of mass of the galaxy \citep{Combes2019}.

In the case of NGC~1068 the brightest molecular line emission extends roughly 15~pc along Position Angle (PA)~114\degr{}, close to the major axis of the \water{} megamaser disc \citep{2016ApJ...823L..12G,2016ApJ...829L...7G,2018ApJ...853L..25I}. The kinematics of the brightest emission is consistent with counter-rotation relative to the \water{} megamaser disc \citep{2019ApJ...884L..28I,2020ApJ...902...99I}, or a tornado-like rotating outflow \citep{2019A&A...632A..61G,2023MNRAS.518..742B}. A strong absorption line is produced by \rottrans{HCN}{3}{2} against the nuclear continuum source. Although the strongest absorption is centred near the systemic velocity (1130 km/s, optical convention, LSRK reference frame, \cite{GG97}, \cite{Gallimore2023}), a blue wing on the absorption line suggests line-of-sight outflow speeds approaching 500~\kms{} \citep{2019ApJ...884L..28I}. Providing additional evidence for nuclear-driven molecular outflow, the high-velocity wings of \rottrans{CO}{6}{5} emission are produced by molecular gas extending a few pc from the central engine along an axis roughly perpendicular to the plane of the \water{} megamaser disc \citep{2016ApJ...829L...7G}. Such outflow might be connected to the one found at larger scales by \cite{GarciaBurillo2014} through clear signatures in the CND, in terms of the kinematics, outflow rates, and enhanced molecular species.  

While ALMA has resolved and begun to explore the kinematics of the obscuring region of NGC~1068, key questions remain. For example, the origin of the apparent counter-rotation is unclear. Furthermore, the molecular outflow is not well-resolved from the high-velocity \water{} megamaser disc. Towards addressing these questions, we used ALMA to observe \rottrans{CO}{3}{2} and \rottrans{HCO$^+$}{4}{3} on the longest available baselines, resulting in 16~mas ($\sim$1.1~pc) angular resolution. These transitions probe the molecular gas with densities between $n(H_2)=10^3$--$10^7\ cm^{-3}$ and temperatures of 50--150 K.

\section{Observations, calibration, and data reduction}

We observed NGC 1068 on 3--4 September 2021 with ALMA using the longest baseline configuration available for Band~7 (C43-9/10). The observations were tuned to observe \rottrans{CO}{3}{2}, \rottrans{HCO$^+$}{4}{3}, and 350~GHz continuum (project 2019.1.01540.S,  PI: P. van der Werf).

The data were calibrated in CASA 6.4.1 \citep{2022PASP..134k4501C} using the standard ALMA pipeline. The pipeline calibrations include bandpass corrections and flux scale calibration based on observations of J0432$-$0210 (3.6~Jy at $\nu = 345$~GHz) and phase-referencing based primarily on observations of J0239-0234 located 2\fdg{}7 from \ngc{1068}. The pipeline calibration was applied independently to three separate observing blocks. After calibration, the separate data sets were merged to produce a single data set for further processing (CASA task {\tt concatvis} with frequency tolerance $=0.6$~MHz). 

The data were averaged in frequency and time to reduce the data volume. The final channel widths are 8958~kHz ($\sim$ 7.53~\kms{}), and the time averaging was 40 seconds. 
For imaging purposes, the visibility data were phase-shifted from Right Ascension RA(J2000) = 02h~42m~40\fs770900, and Declination DEC(J2000) = $-$00\degr~00\arcmin~47\farcs84000 to the position of the nuclear radio continuum source S1, RA(J2000) = 02h~42m~40\fs70905, DEC(J2000) = $-$00\degr~00\arcmin~47\farcs945 \citep{2004ApJ...613..794G}. 
The angular resolution is 16~mas ($\sim$1.1~pc). 
\subsection{Continuum and self-calibration}\label{sec:cont}

\begin{figure*}[h]
\centering
\includegraphics[width=\hsize]{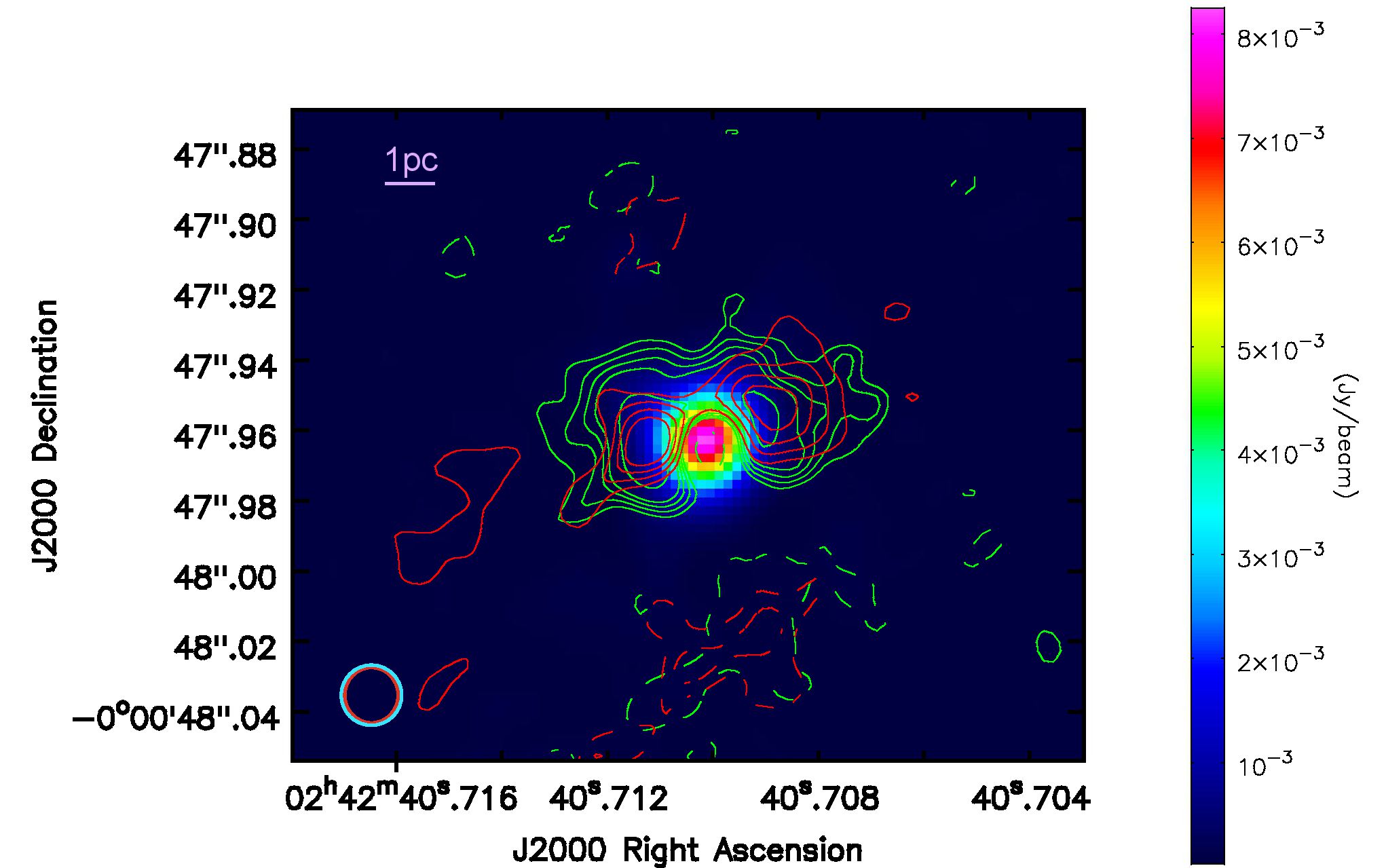} 
\caption{Contours of the integrated \rottrans{CO}{3}{2} (red contours) and \rottrans{HCO$^+$}{4}{3} (green contours) emission overplotted on the 350~GHz continuum image (pseudocolor). The levels for the contours are -4, -3, 3, 4, 5 and 6 times the Root Mean Square (rms) of the \rottrans{CO}{3}{2} moment 0 map (0.024 Jy/beam km/s) and -3, 3, 4, 5, 6 and 9 times the rms of the \rottrans{HCO$^+$}{4}{3} moment 0 map (0.021 Jy/beam km/s). The dashed lines indicate negative contours. The negative features north and south of the AGN are caused by the missing short baselines in the imaging process. The beam size and shape ($\sim$15.5 $\times\sim$16.5 mas with a PA of -43$^\circ$) for the two lines and the continuum is very similar, as marked by the ovals in red and cyan, respectively,} in the lower left corner of the figure. 

\label{Fig:overplot}
\end{figure*}

We used standard self-calibration techniques to improve the final image fidelity of the continuum images \citep{10.1117/12.958828,1981MNRAS.196.1067C}. Self-calibration involved first producing a CLEAN model for the continuum based on line-free channels (CASA task {\tt tclean}) and then using that model to guide phase corrections (CASA tasks {\tt gaincal} and {\tt applycal}). Because the brightest continuum source, S1, is fairly weak (peak flux density $\sim 8$~\mJybeam{}), the self-calibration solutions had to be time-averaged to ensure good solutions. To this end, we repeated the initial round of self-calibration trying a range of discrete time averaging intervals. We adopted an averaging time of 40~s because it produced the lowest failure rate of gain solutions. In the end, we performed 1 round of phase-only self-calibration. We show in Fig.~\ref{Fig:overplot} the resulting continuum image in pseudocolor.

\subsection{Spectral line cubes}
Before imaging, the spectral line data were continuum-subtracted using line-free channels within each spectral window (CASA task {\tt uvcontsub}); the five channels on either end of a spectral window were also excluded from the continuum subtraction. To reduce the data volume further, the continuum-subtracted data were time-averaged by 300 s.  

The \rottrans{CO}{3}{2} line cubes include significant emission from the 100~pc scale `circumnuclear disc' (CND), especially the eastern knot, characterized by strong dust continuum and molecular line emission \citep{GarciaBurillo2014} \footnote{This area was first defined in \cite{Viti2014} (see Fig. 1) and has continued to be used in subsequent studies, such as \cite{GarciaBurillo2016} and \cite{GarciaBurillo2019}.}. Therefore, for each line channel, we imaged a large, 9\arcsec{} field around S1 and deconvolved using CLEAN (CASA task {\tt tclean}). We used Briggs weighting with the robust parameter set to 2.0 \citep{1995PhDT.......238B} and 2.5~mas pixels. Initial line images showed a strong ripple pattern produced by undersampled, large-scale emission. We therefore flagged baselines shorter than $1200~\mbox{k}\lambda$ (structures larger than about 0\farcs2) to reduce the ripple artefacts on the final line images. 

We used the same approach for the images of \rottrans{HCO$^+$}{4}{3}. However, there was no significant contribution from the CND, so we imaged only the central 0\farcs75. The integrated \rottrans{CO}{3}{2} and \rottrans{HCO$^+$}{4}{3} emissions are included in Fig.~\ref{Fig:overplot}.

\section{Spectral line analysis} \label{sec:gaussianFits}

To understand the kinematics of the gas close to the AGN, we analysed the central 250~mas (17.5~pc) by fitting a single Gaussian profile to the spectrum at each pixel of the line cubes. This approach provides a clearer representation of the velocity structure, as the more conventional moment maps yield representations that are harder to interpret due to the low signal-to-noise ratio of the cubes. The Gaussian profile is expressed as,
\begin{equation}\label{Gaussianprofile}
    S_{\nu}(v) = A \exp{\left[-(v - v_0)^2/2s^2\right]},
\end{equation}
where $S_{\nu}(v)$ is the observed flux density as a function of recessional velocity $v$,  A is the line amplitude, $v_0$ is the velocity at the line centre, and $s$ is the linewidth (the standard deviation in the definition of the Gaussian function), which includes velocity dispersion but also contributions from rotation and radial motions within the beam. We constrained the fits in velocity to 900~\kms{}$ < v_0 < 1350$~\kms{} resulting in fitted mean velocities of $v_0 <300$~\kms{}. This was done to capture the main kinematic component near the AGN's systemic velocity. As a result, any high-velocity features outside this range, whether redshifted or blueshifted, are not accounted for in these fits. Additionally, we constrained the amplitudes to $|A| > 1.5\times\mbox{rms}$, where rms is the background noise of the line cubes, and the linewidths to 30~\kms{}$ \leq s \leq 190$~\kms{}.  
We rejected spectra for which the integrated signal-to-noise ratio was less than 3. As a consequence, only fits in emission were kept. 
Further analysis of the central beam spectra and the Position-Velocity (PV) diagrams reveal high-velocity absorption features at around  $v_0\sim$400~\kms{}. These trace a distinct physical process and are discussed separately in Sects. \ref{sec:pvDataandG} and \ref{sec:agnspectra}.

We performed a careful visual inspection of each of the spectral line fits in emission and concluded that the Gaussian model represents the line spectra well. However, in a small number of spectra closest to S1, where the lines are the broadest, we see evidence of multiple spectral peaks. 
This Gaussian fitting analysis effectively compresses the spectral line cubes into sky maps of the parameters $A$, $v_0$, and $s$. We used these cubes to further analyse the velocities, the linewidths and the morphology of the emitting regions in Sects. \ref{sec:velslinewidthG} and \ref{sec:pvDataandG}.

\section{Results}\label{sec:resultsdiscussion}
\subsection{Integrated line emission}

Figure~\ref{Fig:overplot} shows the integrated \rottrans{CO}{3}{2} and \rottrans{HCO$^+$}{4}{3} images as contours over the 365~GHz continuum image in pseudocolor. The continuum is dominated by a compact source with peak intensity 8.3~mJy\,beam\mone{}. In our ALMA image, the continuum peak is offset by $\sim$25~mas northwest of the VLBI position of S1. This is consistent with typical calibration errors in the ALMA calibration system \citep{2016ApJ...829L...7G}. Perhaps the offset is real, but it seems unlikely. The integrated flux density of the continuum source is consistent with an extrapolation of the cm-wave flux densities, suggesting that the compact mm-wave continuum source is identically S1. We speculate that the offset might result from differences in the cm-wave and mm-wave structure of the phase reference, J$0239-02$. Going forwards, we assume that the compact, 365~GHz continuum source coincides with the S1 VLBI continuum source and adjust the coordinates accordingly. We note that a similar position offset was observed in previous ALMA observations of the same galaxy \citep{2016ApJ...829L...7G}, and more recently in the case of the Circinus galaxy \citep{2022A&A...664A.142T}.

The integrated CO and the HCO$^+$ images show a morphology roughly comparable to that found in previous ALMA observations (\cite{2016ApJ...823L..12G, 2016ApJ...829L...7G, 2019A&A...632A..61G}, with angular resolutions of 4, 5.9 and 2--6 pc, respectively). Any differences can be largely attributed to the application of our self-calibration procedure and the deliberate omission of the short baselines in our data analysis.
The line emission resolves into two bright peaks bracketing the nuclear continuum source and oriented nearly east-west (PA~102\degr{}). The integrated line peaks are separated by roughly $\sim30$~mas ($\sim 2$~pc). North of the continuum source, marginally resolved line emission forms a bridge between the two emission line peaks, but there is a conspicuous gap south of the continuum source. Weak HCO$^+$ emission is detected up to $\sim 50$~mas (3.5~pc) from the nucleus, towards the northwest direction, and we also find weak HCO$^+$ line absorption against the nucleus (See Fig. \ref{Fig:overplot}). The CO emission seems to extend towards the direction of the bright eastern knot, since there is a weak and smaller source, $\sim 65$~mas (4.6~pc) to the southeastern direction (located at 2:42:40.7153 in RA and -0:00:47.9934 in DEC). While the morphology of the emission line sources are similar and follow the position angle close to that of the central radio source S1 ($104\degr{}$) defined by \cite{2004ApJ...613..794G}, their position angles differ from that of the \water{} maser disc by $\sim$12 \degr{} (c.f. Fig. \ref{Fig:RADECMeanVel}), and the HCO$^+$ seems to come from a geometrically thicker disc, when comparing it to that of the CO.
We also see that cutting the short baselines during our cleaning process created negative bowls in both line maps $\sim$ 50 mas north and south from the AGN.

\subsubsection{The \rottrans{CO}{3}{2} to \rottrans{HCO$^+$}{4}{3} line ratios}\label{sec:lineratios}
\begin{figure}
   \centering\vspace{-0.5cm}
   \includegraphics[scale=0.5]{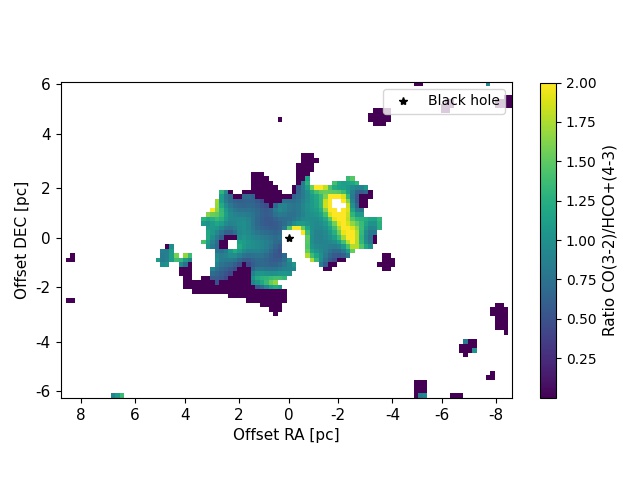}\vspace{-0.5cm}
  \caption[width=0.55\hsize]{Ratio of the integrated flux of \rottrans {CO}{3}{2} to \rottrans{HCO$^+$}{4}{3} as a function of position. Areas where the signal to noise ratio of the CO line are $<7$ rms have been suppressed. The central black star marks the position of the black hole.}
  \label{Fig:lineratios}
  \end{figure}

Figure~\ref{Fig:overplot} shows that the morphology of the compact molecular discs derived from the velocity-integrated line fluxes of the \rottrans{HCO$^+$}{4}{3} and \rottrans{CO}{3}{2} are approximately equal with some spatial variations. 
Figure \ref{Fig:lineratios} displays the line ratios quantitatively across the source and shows a systematic behaviour being $\sim 1$ near the main major axis of the molecular system (considering a position angle of 114$^{\circ}$), and decreasing to $\sim 0.5$ above and below this axis. A prominent feature in this image is a strip with higher values $\sim$ 2 pc to the west from the AGN. It is interesting to note that this feature coincides with the low velocity region of the gas related the drop to almost 0 km/s velocities in the west side of the disc (See Sect. \ref{sec:velslinewidthG} and Fig. \ref{Fig:RADECMeanVel}). To enhance the areas with the colour scale we have limited the values displayed to a maximum of 2.0, nevertheless the few pixels that were left in blank in the middle of this strip reach higher values: in average the ratio of the lines in those pixels is of 4.
\cite{Viti2014} studied the entire circumnuclear disc (CND) of NGC~1068 in many molecular lines, covering a larger area, but at a lower spatial resolution that the data presented here. They  defined four main regions in addition to the AGN: the CND regions, called eastern knot, western knot, CND-north and CND-south. They found much higher values of the CO/HCO$^+$ ratio than those found here, ranging from 9 to 16, 
and 11 at the AGN, all at a spatial resolution of 100 pc compared to the 1.1 pc resolution of the current data. They found the starburst ring (SB ring) to have a much higher line ratio of 72 at 400 pc resolution. 
We discuss the interpretation of these ratios further in Sect. \ref{sec:lineratiodiscussion}.

\subsection{Recessional velocities and linewidths} \label{sec:velslinewidthG}

\begin{figure*}
   \centering
   \includegraphics[width=1\hsize]{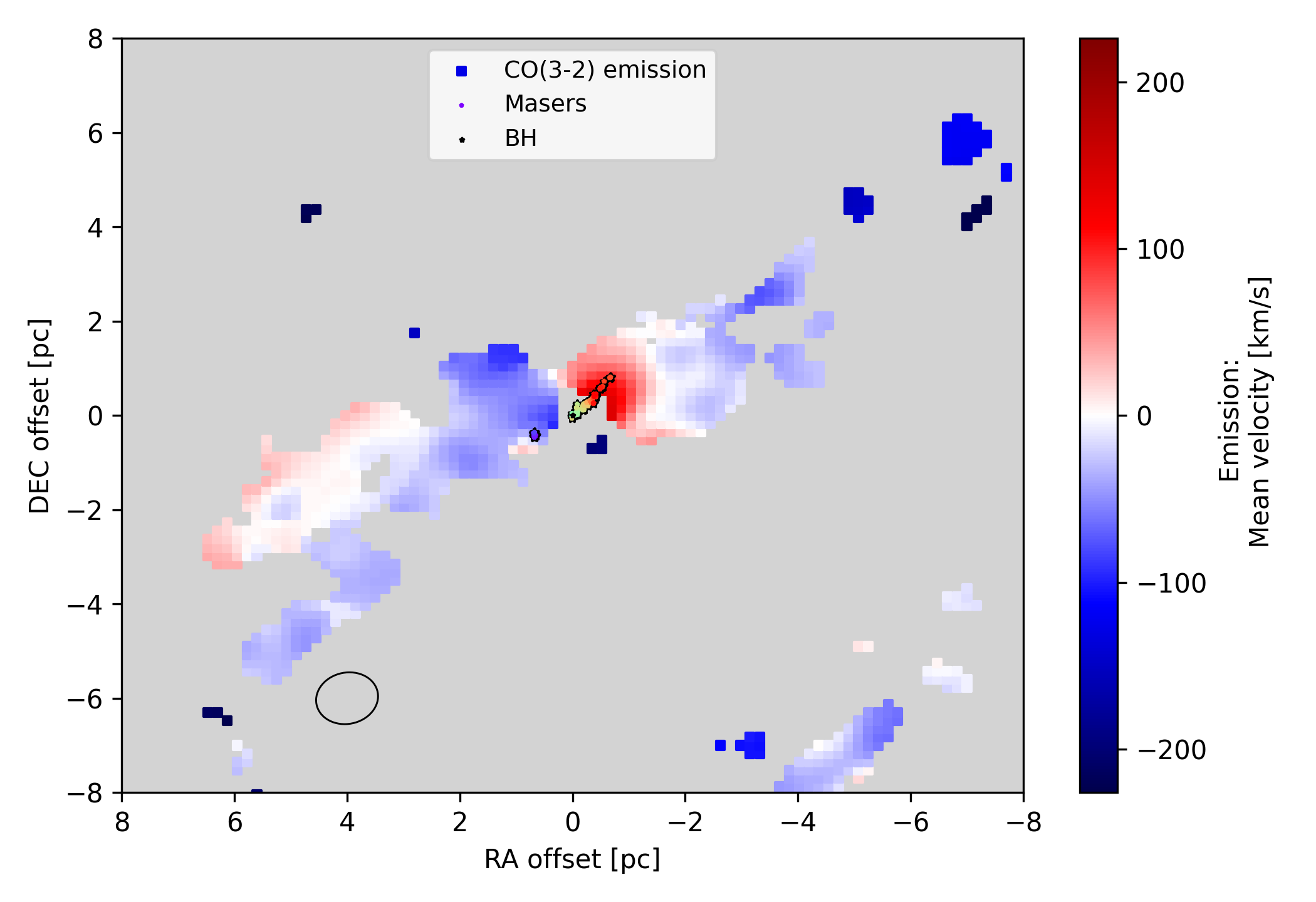} \vspace{-1.25cm}
   
   \includegraphics[width=1.02\hsize]{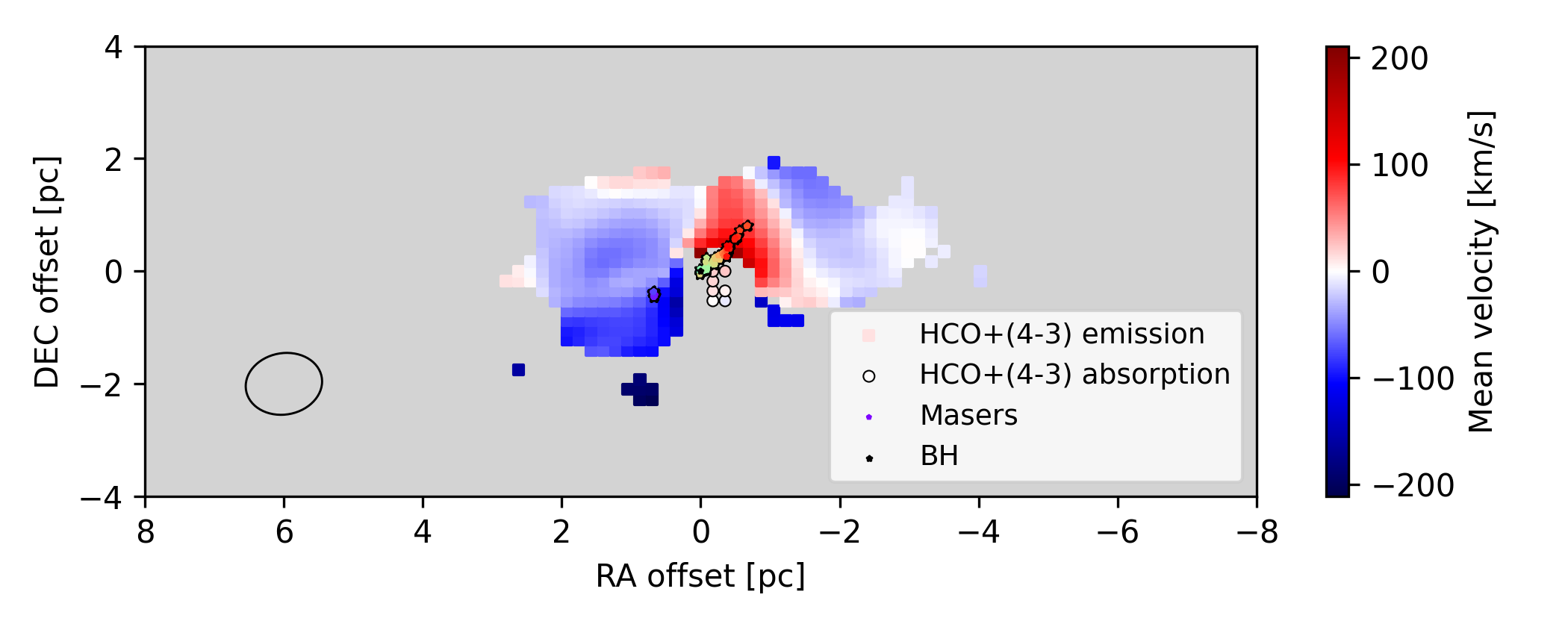}  
    \caption[width=1\hsize]{Maps of CO and HCO$^+$ recessional velocities in the nuclear region of NGC~1068. Velocities relative to systemic are shown in blue--red pseudocolor according to the colorbar at right. Top: \rottrans{CO}{3}{2} emission velocities. Bottom: \rottrans{HCO$^+$}{4}{3} emission and absorption velocities. In each plot, the ellipse in the lower left corner represents the beam size. North is up and east is left. The black star marks the position of the supermassive black hole and the small colourful dots represent the \water{} megamaser disc \citep{2004ApJ...613..794G}.} \label{Fig:RADECMeanVel}
\end{figure*}

\begin{figure*}
   \centering
   \includegraphics[width=1\hsize]{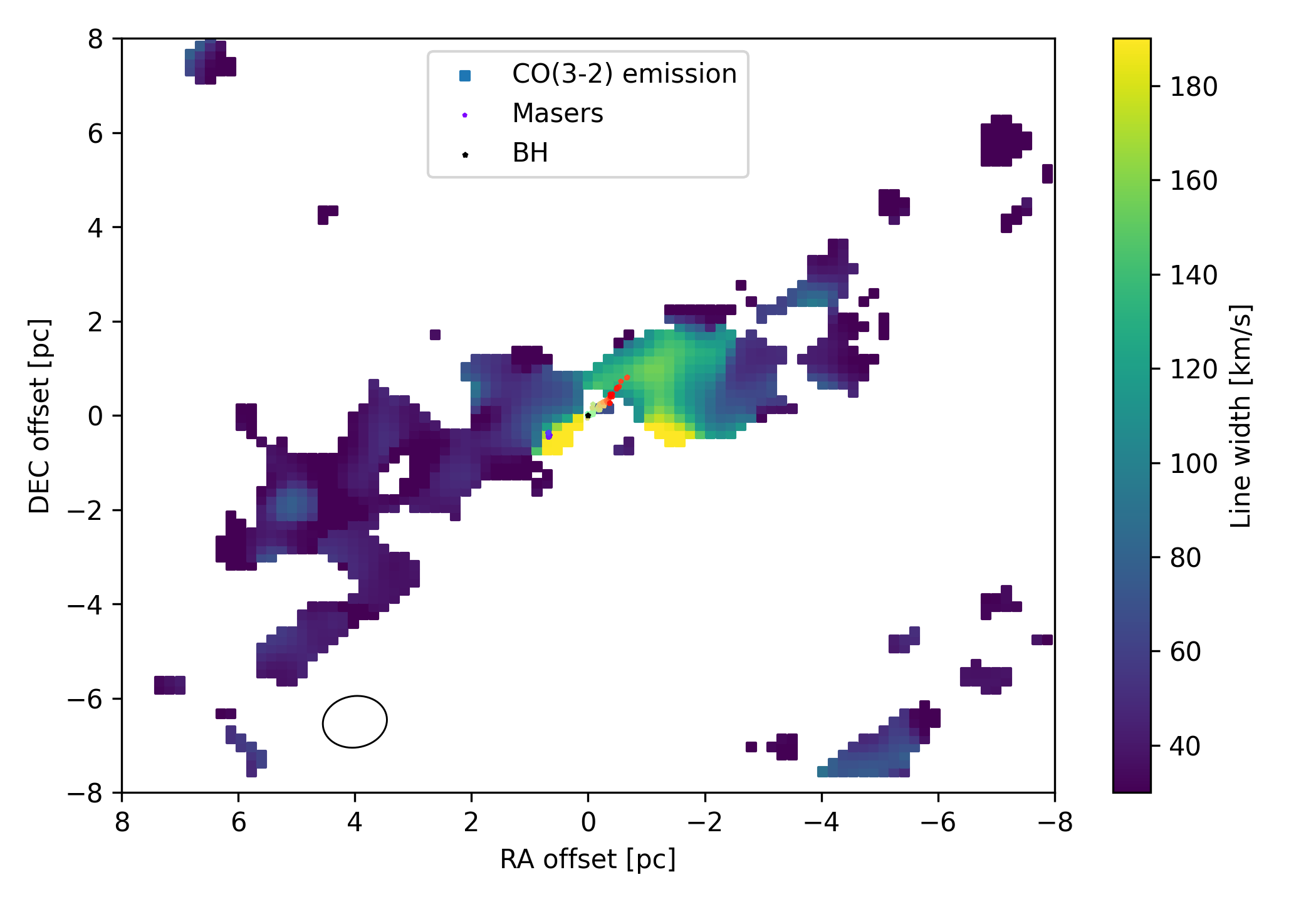} 
      \includegraphics[width=1.01\hsize]{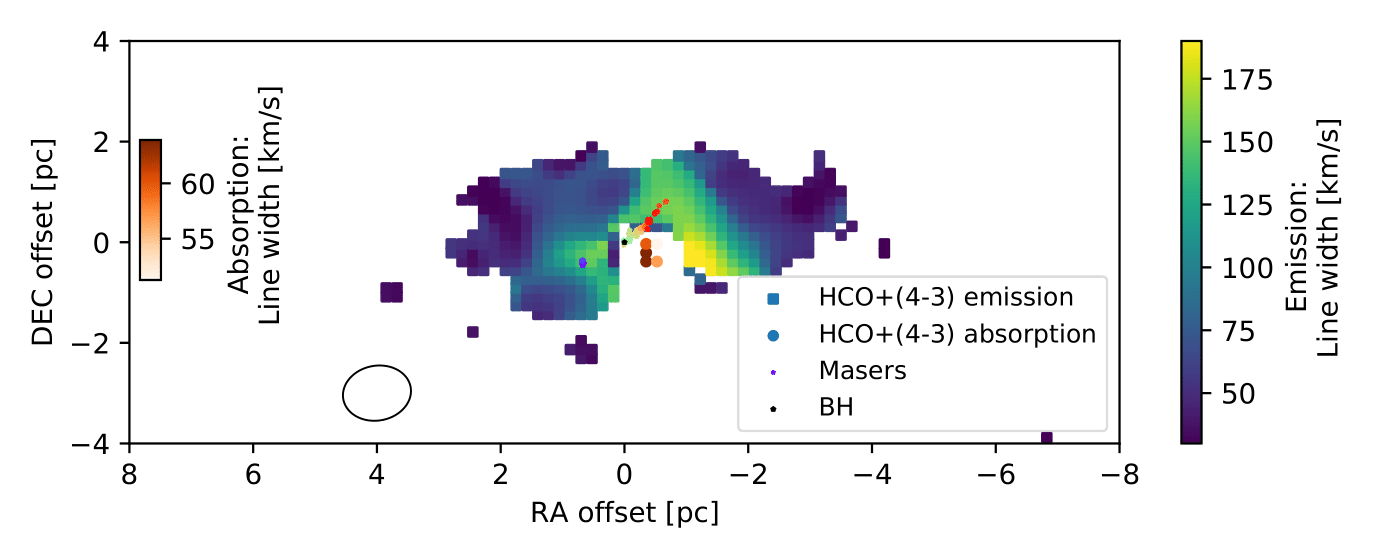}  
   \caption[width=1\hsize]{Maps of CO and HCO$^+$ linewidths $s$ (Eq. \ref{Gaussianprofile}) in the nuclear region of NGC~1068. Emission linewidths are shown in blue--yellow pseudocolor according to the colorbar at right. Top: \rottrans{CO}{3}{2} emission linewidths. Bottom: \rottrans{HCO$^+$}{4}{3} emission and absorption linewidths (white -- brown colorbar shown in the inset). In each plot, the ellipse in the lower left corner represents the beam size. North is up and east is left. The black star marks the position of the supermassive black hole and the small colourful dots represent the \water{} megamaser disc \citep{2004ApJ...613..794G}.} \label{Fig:RADECVelDisp}
\end{figure*}

We created velocity and linewidth maps derived from Gaussian fits to the cube spectra (Sec.~\ref{sec:gaussianFits}). The results are shown in Figs.~\ref{Fig:RADECMeanVel} and \ref{Fig:RADECVelDisp}. It is important to note that the fit was done with the aim of representing the best kinematic system centred at the AGN, which resulted in Gaussians with mean velocities between -230 and 320 km/s. Therefore any high-velocity component, either redshifted or blueshifted, is not represented in these plots. Recessional velocities are predominantly redshifted (blueshifted) immediately west (east) of S1, consistent with the sense of rotation of the water megamaser disc \citep{2004ApJ...613..794G}. 
The velocity pattern reverses beyond about 2~pc east and west of S1, consistent with the apparent counterrotation reported by \cite{2019ApJ...884L..28I} and \cite{2020ApJ...902...99I}.  However, there is also an extended region of blueshifted CO emission extending to about 6~pc southeast of S1.  

Not surprisingly, linewidths are highest near the central continuum source. In particular, there are wings of enhanced linewidth along the major axis of the \water{} megamaser disc \citep{2004ApJ...613..794G}. The highest linewidths, approaching $s = 200$~\kms{}, are found about 1~pc southwest of the radio continuum source. We further discuss this in Sect. \ref{sec:lbG}. Beyond about 2--3~pc from S1, linewidths fall to $s = 40$--60~\kms{}. There is also a region of low linewidth about 1~pc northeast of the nucleus; the velocities in this region tend to be slightly blueshifted relative to systemic in both CO and HCO$^+$.

\begin{figure*}
   \centering
   \includegraphics[width=1\hsize]{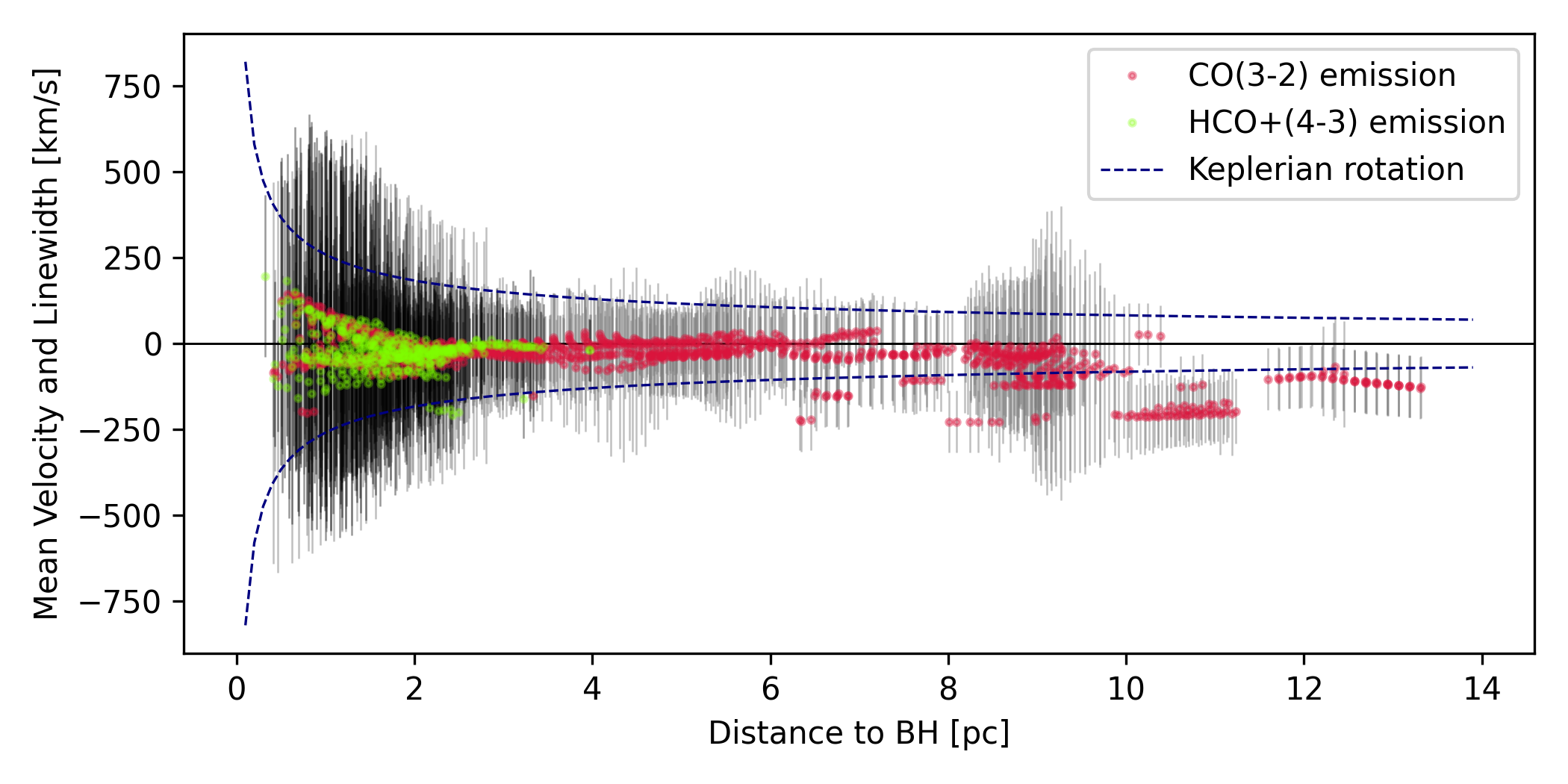}
    \caption[width=1\hsize]{Mean velocities obtained from the Gaussian fits to the \rottrans{CO}{3}{2} (red) and \rottrans{HCO$^+$}{4}{3} (green) emission, with bars representing best-fit linewidths ($3\times s$, Eq. \ref{Gaussianprofile}), plotted against the distance to the central continuum source or S1.} \label{Fig:VelDispDist} 
   \end{figure*}

Figure~\ref{Fig:VelDispDist} shows the best-fit velocity centroids and linewidths as a function of distance from the S1 continuum source. Two results stand out. First, within about 1~pc of the central continuum source or S1, the red- and blueshifted mean velocities in both lines (red dots for \rottrans{CO}{3}{2} and green dots for \rottrans{HCO$^+$}{4}{3}) are asymmetric with respect to the systemic velocity (marked by 0 in this plot); the mean velocities of the western redshifted region (upper half) reach $v_0 \sim 200$~\kms{}, but the eastern blueshifted region  (lower half) shows velocities typically below $v_0 = 100$~\kms{}. Second, beyond about 2~pc from S1, the mean velocities are typically within a linewidth of the systemic velocity, well below the rotation curve extrapolated from the \water{} megamaser disc (dashed curves, cf. \cite{2019ApJ...884L..28I, Gallimore2023}). Overall, the pattern suggests a switch in the molecular gas kinematics: inside about 2~pc, the velocities are broadly consistent with rotation, but outside the kinematics appear to be dominated by random motions, since the mean velocities are very low, of the order of 30 km/s in average, but the velocity dispersion remains constant with radius and of the order of 50 km/s.

\subsubsection{Spectra at the AGN position}
\label{sec:agnspectra}

\begin{figure*}
   \centering
   \includegraphics[width=0.45\hsize]{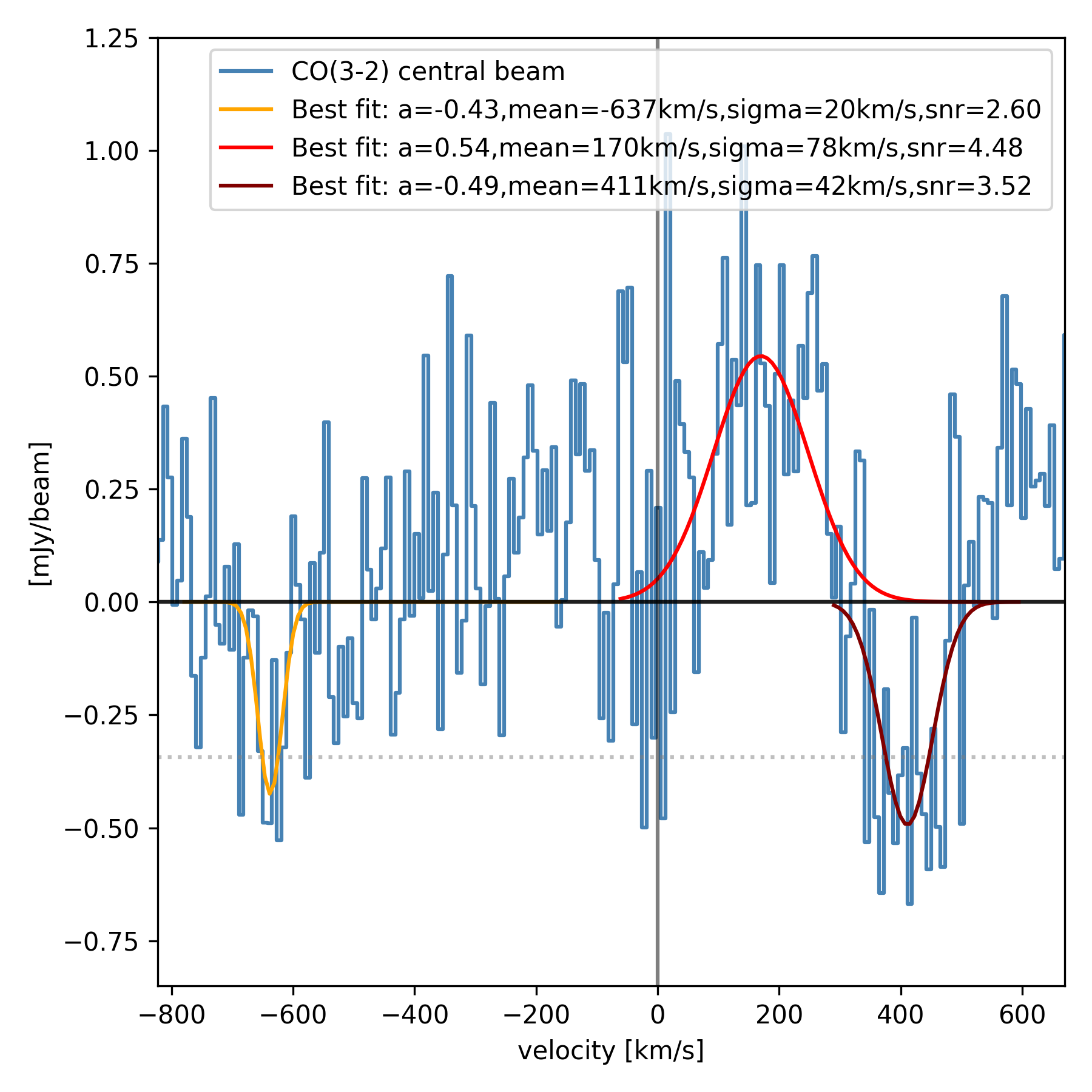} \includegraphics[width=0.45\hsize] 
{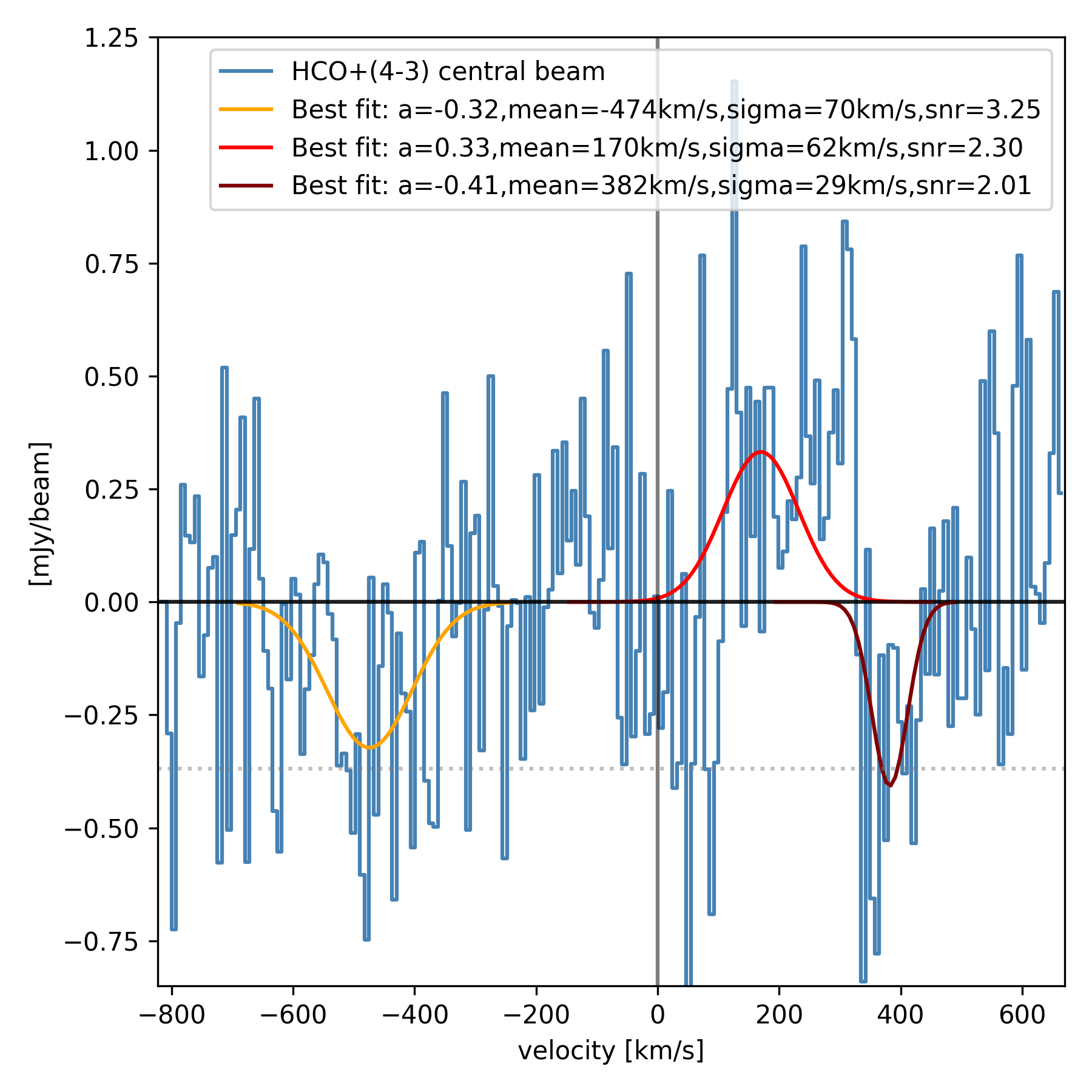} 
   \caption{Gaussian fits of the absorption and emission features of the \rottrans{CO}{3}{2} and \rottrans{HCO$^+$}{4}{3} spectra belonging to the central beam on top of the peak of the continuum relative to the systemic velocity. The dotted line marks -1 times the rms of the spectrum. We used a value for the systemic velocity of 1130 km/s. } \label{Fig:abs}
   \end{figure*}

Figure~\ref{Fig:overplot} shows absorption in the \rottrans{HCO$^+$}{4}{3} at the position of the continuum peak. In the case of the \rottrans{CO}{3}{2} line there is a hole at the same position and where the absorption is seen for the other line. The area marked by the dashed green contour has the shape of an ellipse which centre is displaced with respect to the peak of the continuum by 1 or 2 pixels. This same region is characterized by the absence of emission of the \rottrans{CO}{3}{2} transition. To analyse this in more detail we extracted the spectra of both lines, using the CASA task viewer, integrated over one beam centred at the position of the AGN, this is, on top of the peak of the continuum (See Fig. \ref{Fig:abs}). To our surprise, in the spectra of the \rottrans{CO}{3}{2} we see several features in absorption as well. In this section we characterise the profiles of both molecular transitions by fitting Gaussians to their spectra and calculating their significance.

In the case of the \rottrans{CO}{3}{2} transition we find high-velocity absorption centred at 
$\sim 411$ km/s with a signal to noise ratio (S/N) of 3.5. The other possible absorption feature at 
$\sim -637$ km/s does not present a S/N high enough to be considered a real detection (see Fig. \ref{Fig:abs} top).

The \rottrans{HCO$^+$}{4}{3} spectrum, in contrast, shows absorption centred at 
$\sim -492$ km/s with a Signal-to-Noise Ratio (S/N) of 3.03, while the other possible absorption feature centred at $\sim 359$ km/s does not present a S/N high enough to be considered a real detection (see Fig. \ref{Fig:abs} bottom). 
Both spectra present broad emission features at around +170 km/s that seem to be deformed by the absorption at high velocities, but it is only significant enough in the case of the \rottrans{CO}{3}{2} line, with a S/N of 4.5. While the emission seems to peak at similar velocities with that of the \rottrans{HCN}{3}{2} at 256 GHz found in \cite{2019ApJ...884L..28I} (comparing to the raw spectrum), the strong nuclear absorption found in the same line differs significantly to our data. That is, their line profile is asymmetric, it peaks at the systemic velocity and it displays a blue wing extending to roughly -450 km/s. This HCN absorption was interpreted as a high-velocity nuclear molecular outflow. This feature is not present in our \rottrans{HCO$^+$}{4}{3} nor \rottrans{CO}{3}{2} spectrum.

The 350 GHz continuum flux at the nuclear position is 8.3 mJy or in brightness temperature $T_{c}=550$ K, and the negative fluxes in Fig. \ref{Fig:abs} near +400 km/s represent optical depths of $\sim 0.05$ corresponding to a brightness temperature decrease of $\Delta T_{a}\sim -30$ K. However there is emission at adjacent velocities, and the major-axis PV-diagrams (shown in Fig. \ref{Fig:pvmajco}) suggest that there is actually some emission in both lines at the nuclear position over most velocities between -400 and at least +400 km/s.  Both the emission and absorption features seen in Fig. \ref{Fig:abs} are probably considerably broader than is apparent in the figure, but cancel each other at the velocities where they overlap. By interpolation from adjacent emission in Figs. \ref{Fig:abs} and \ref{Fig:pvmajco} we estimate that in the absence of absorption, the actual line emission at the nuclear position near +400 km/s would be $\sim 0.5$ mJy and the actual absorption optical depths are more likely $\sim 0.1$ in both lines.  

The physical conditions in the gas responsible for the absorption are discussed in Sect. \ref{sec:noncircular}.

\subsection{Position--Velocity diagrams} \label{sec:pvDataandG}
\begin{figure*}
   \centering
  
   \includegraphics[scale=0.5]{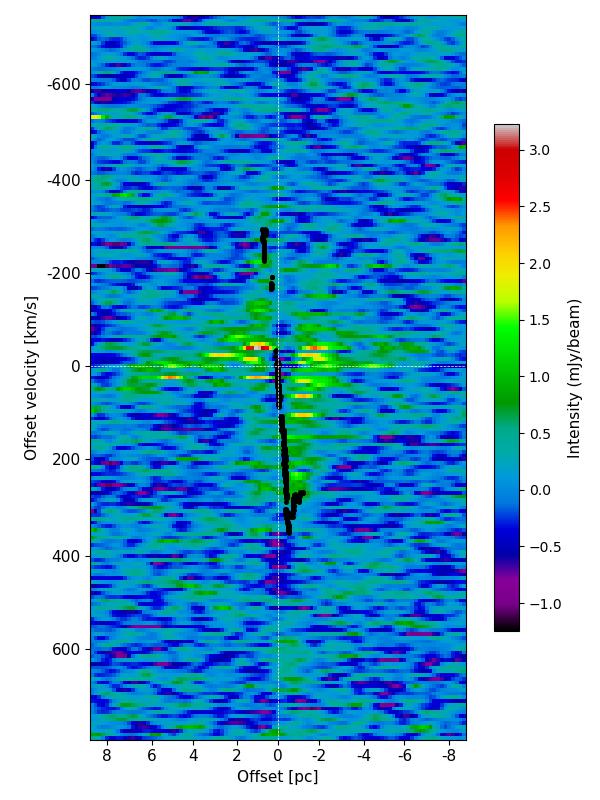}\includegraphics[scale=0.5]{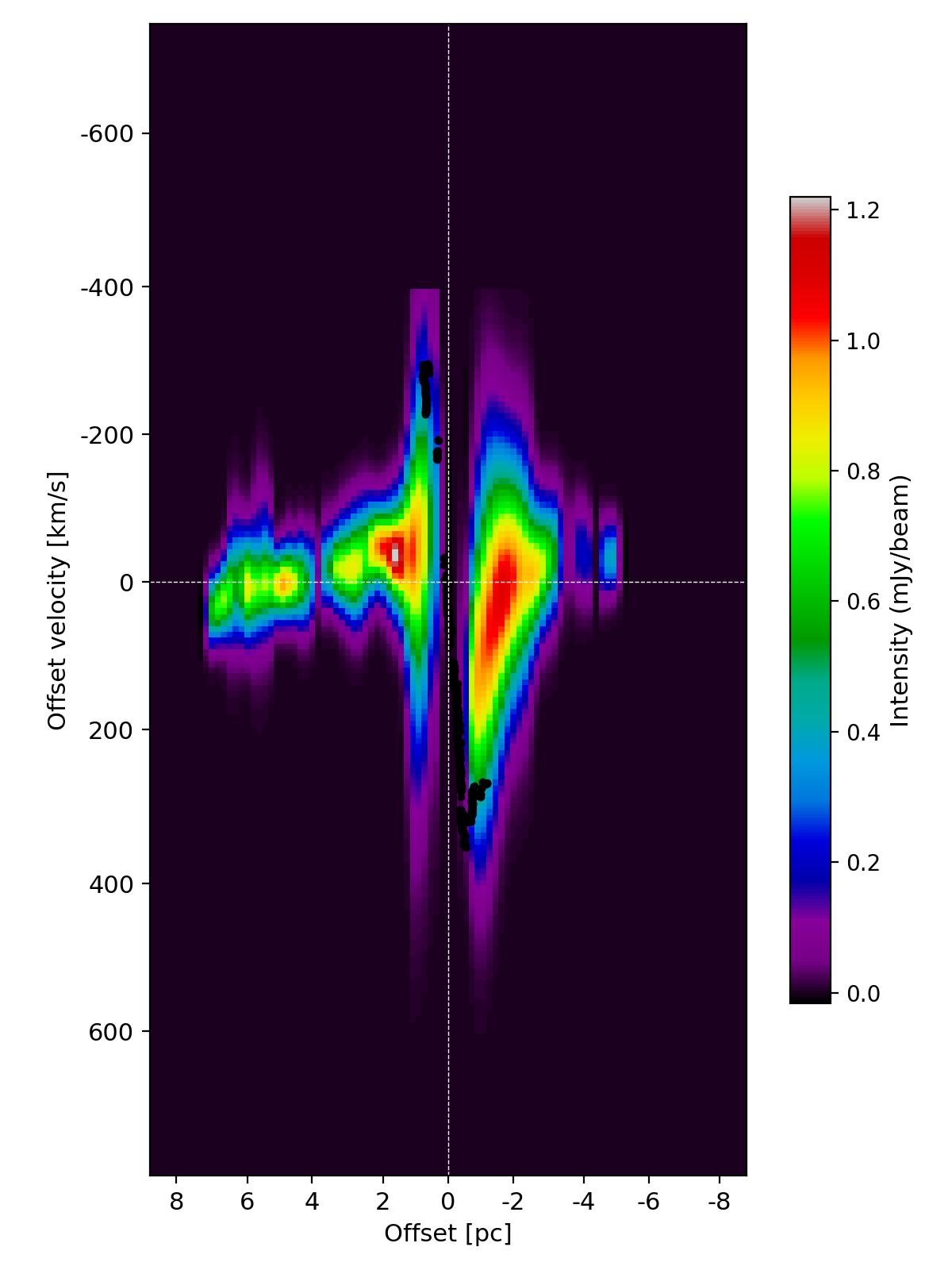} 
   \hspace{0.1cm}\includegraphics[scale=0.5]{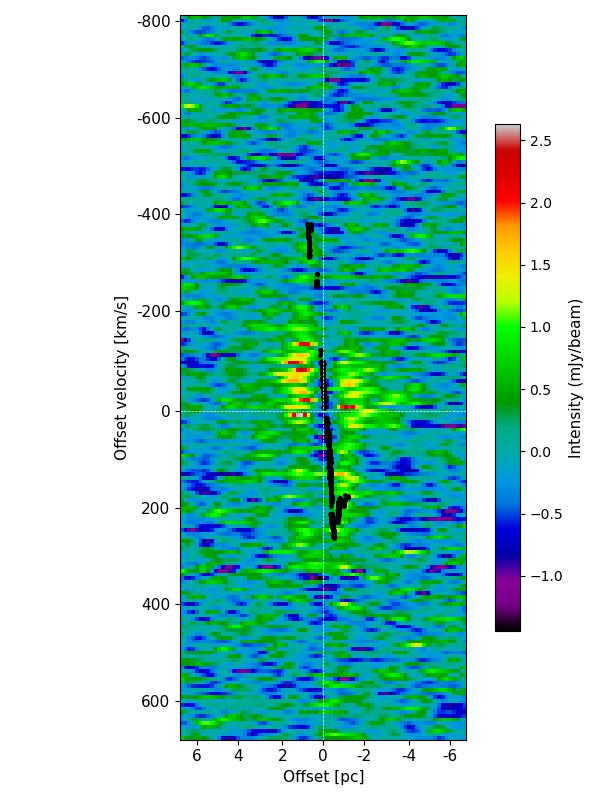}\hspace{0.18cm}\includegraphics[scale=0.5]{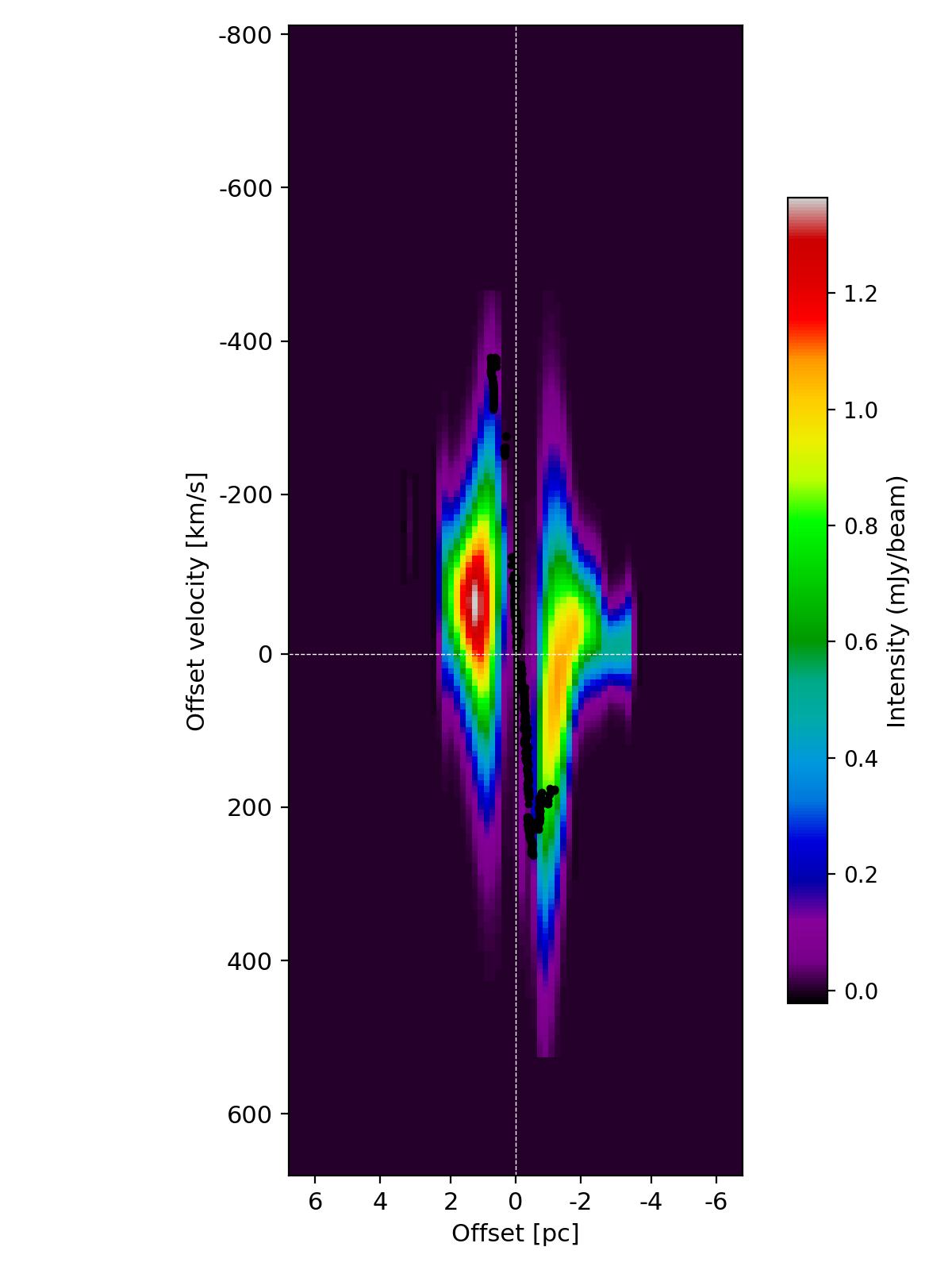}

   \caption[width=0.55\hsize]{Position-velocity diagram of the \rottrans{CO}{3}{2} (top) and the \rottrans{HCO$^+$}{4}{3} (bottom) along the major axis originated from the data (left) and the Gaussian fits (right). The black dots on the two top panels mark the positions of the \water{} masers \citep{Gallimore2023}. East is almost to the left taking into account the PA used to make the diagram (PA=114$^{\circ}$).} \label{Fig:pvmajco} 
   \end{figure*}

Figure~\ref{Fig:pvmajco} (left) shows the position-velocity (PV-) diagrams along the major axis of the \rottrans{CO}{3}{2} and the \rottrans{HCO$^+$}{4}{3} using a PA of 114$^{\circ}$. 
We also plot in Fig. \ref{Fig:pvmajco} (right) the analogue PV-diagrams obtained from the Gaussian models. 
The diagrams for both lines are very similar. They show a lot of asymmetry between the west and east sides of the disc. While the west side has a large area with high intensities the east side presents the peak intensity in a smaller area. The peaks are symmetric with radius, but not with velocity and they differ in flux. The mean velocities decay quickly in the inner $\sim$2 pc in both sides of the disc, nevertheless the western side has a strong signal of apparent counter-rotation. This is, that there is signal in the quadrants of the PV-diagrams where there should not be if we had a kinematic system rotating purely in one sense. The linewidths are of the order of 200 km/s close to the nucleus and get abruptly reduced to around 50 km/s at r>2 pc.  
The Gaussian models are meant to only represent the emission, but it is clear that there are some areas with absorption features in the PV-maps of the data (left). The missing emission at systemic velocity coinciding with the AGN stands out clearly in both lines, but the absorption at high velocity (at around 400 km/s) is very prominent in the \rottrans{CO}{3}{2} map.

\begin{figure*}
   \centering   
    \includegraphics[scale=0.9]{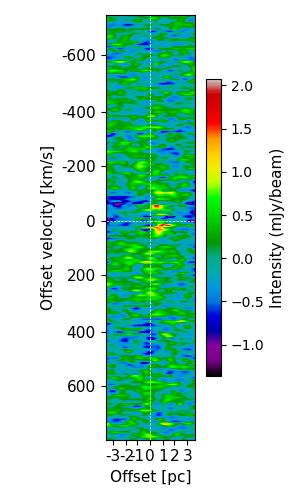}\includegraphics[scale=0.9]{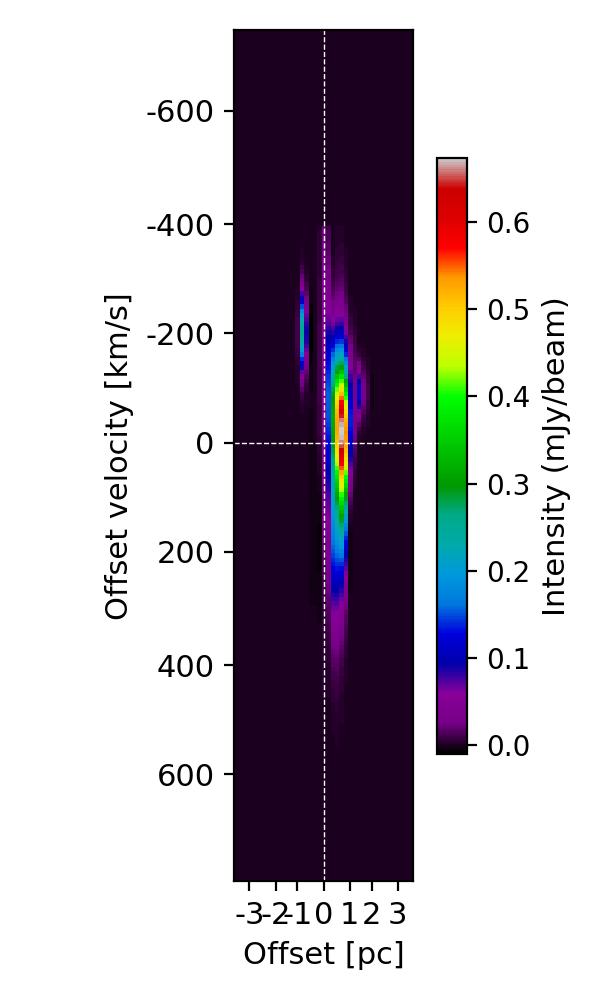}\vspace{-0.9cm}
    
    \includegraphics[scale=0.9]{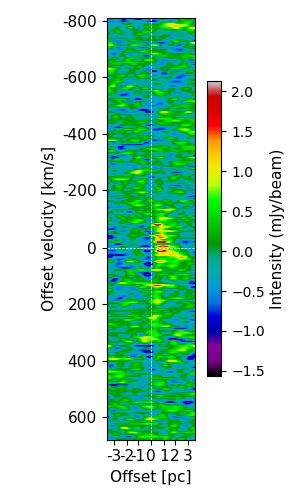}\includegraphics[scale=0.9]{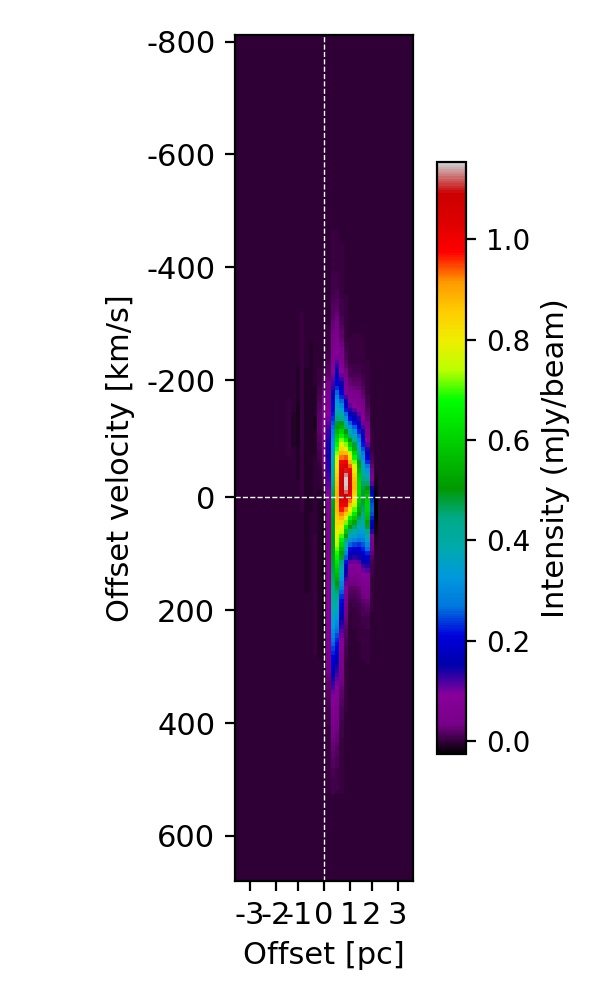}

  \caption[width=0.55\hsize]{Position-Velocity diagram of the \rottrans{CO}{3}{2} (top) and the \rottrans{HCO$^+$}{4}{3} (bottom) along the minor axis originated from the data (left) and the Gaussian fits (right). North is almost to the right taking into account the PA used to make the diagram (PA=24$^{\circ}$).} \label{Fig:pvminco}
\end{figure*}

Figure \ref{Fig:pvminco} shows the position-velocity diagrams along the minor axis of the \rottrans{CO}{3}{2} (left) and the \rottrans{HCO$^+$}{4}{3} using a PA of 24$^{\circ}$. 
We also plot in Fig. \ref{Fig:pvminco} (right) the analogue PV-diagrams obtained from the Gaussian models for the minor axis. 

\section{Discussion}\label{sec:discussion}
Here we consider further two issues that became evident in exploring the data that are difficult to explain in the context of gravitationally bound, rotating disc models, namely:

\begin{itemize}
    \item The very low observed velocities compared to the predictions of Keplerian rotation about the central black hole.
    \item The east-west asymmetries in the PV-diagrams very close to the active nucleus.
\end{itemize} 

For this, we first show that the large linewidth values can  arise from a beam smearing effect with the help of the tool 3DBarolo \citep{2015MNRAS.451.3021D}, and then we test the possibility of having asymmetric drift assuming that the beam smearing is not causing the large velocity dispersions.

\subsection{Line broadening}\label{sec:lbG}
The high linewidths and low rotational speeds observed suggest that the molecular gas might be highly turbulent within the central $\sim 2$~pc. On the other hand, the 16~mas (1.1~pc) beam does not resolve the molecular disc traced by \water{} megamaser emission. As rotational speeds reach $\sim 330$~\kms{} in the maser disc, 
the beam-averaged linewidth is expected to be large. We therefore used 3DBarolo \citep{2015MNRAS.451.3021D} to simulate an ALMA observation of the \water{} megamaser disc and assess the contribution of rotational kinematics to the observed linewidth. The software package 3DBarolo allows to model the rotation curves and other parameters of the disc. It uses a tilted ring model, which consists of nested rings with constant circular velocities each allowing for peculiar morphologies as warped or lopsided systems, and considers the beam smearing effects by adding a convolution step before providing the final model. 
We set up the simulation using the observed rotation curve and orientation of the \water{} megamaser disc \citep{Gallimore2023}; the parameters are summarised in Table~\ref{tab:baroloSimPar}.  As \water{} masers necessarily trace hotter and denser molecular gas, we chose a larger outer radius (40~mas = 2.8~pc) than the observed megamaser disc to simulate possible emission from CO or HCO$^+$. We also assumed a velocity dispersion $\sigma = 15$~\kms{},
roughly two channels, to ensure Nyquist sampling of the simulated spectra. The true velocity dispersion of the maser disc might be smaller, but this simulated dispersion is much less than the range of rotation speeds averaged over the beam and has little impact on the results. The ring width, $\delta r = 2$~mas was chosen as a compromise between finely sampling the Keplerian rotation curve near the inner radius and computation speed. In sum, the simulation involved 17 nested rings between 5 and 40 mas radius. This set up has the implicit assumption that the kinematics defined by the \water{} masers extend towards larger radii. 

\begin{table}[tbh]
    \caption{3DBarolo simulation parameters}
    \label{tab:baroloSimPar}
    \centering
    \begin{tabular}{lrll}
    \hline\hline
    Parameter & Value & Unit & Description \\
    \hline
        $r_{in}$  & 5 & mas & Inner radius\\
        $r_{out}$ & 40 & mas & Outer radius \\
        $v_0$     & 404 & \kms{} & Keplerian rotation speed \\
        $r_0$     & 6.7 & mas & Radius for $v_0$ \\
        $v_{sys}$ & 1130 & \kms{} & Systemic velocity \\
        $\sigma$ & 15 & \kms{} & Velocity dispersion \\
        $i$       & 104.5 & degrees & Inclination \\
        $\delta r$ & 2 & mas & Ring width \\
        $z_0$     &   1 & mas & Disc scale height    
    \end{tabular}
\end{table}

We performed two simulations changing only the emissivity profile of the rings. In Model~1, the emissivity is constant as a function of radius, and in Model~2, the emissivity law falls as $r^{-2}$. Models were calculated on 0.5~mas pixel grids (i.e. $5 \times$ narrower pixels than used for the ALMA data). The objective of having smaller pixels is to be able to capture the whole range of velocities observed in the masers, especially the highest velocities closer to the BH that would help to create broader lines by beam smearing. The resulting model cubes were then convolved with the ALMA beam ($16.3 \times 15.1$~mas, PA $-45\degr{}$). Finally, we repeated the Gaussian fitting analysis for the simulated cubes (see Sec.~\ref{sec:gaussianFits}). 

\begin{figure*}
   \centering
   \includegraphics[width=1\hsize]{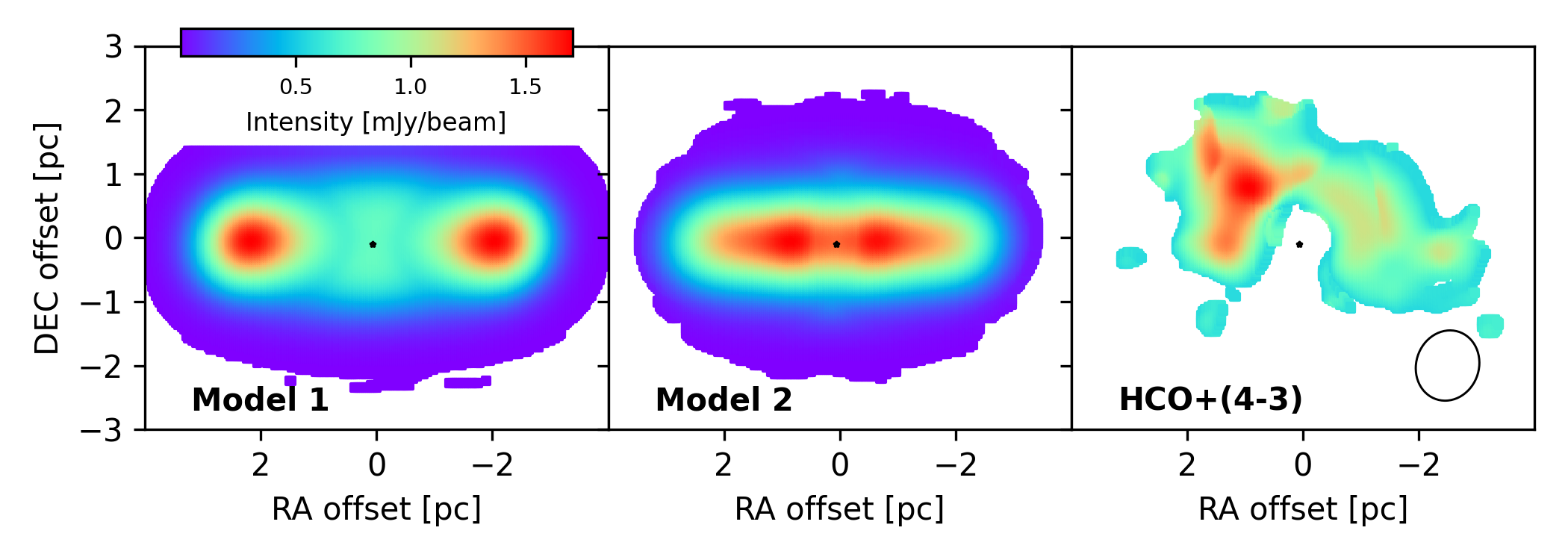}\vspace{-1.4cm}
   \includegraphics[width=1\hsize]{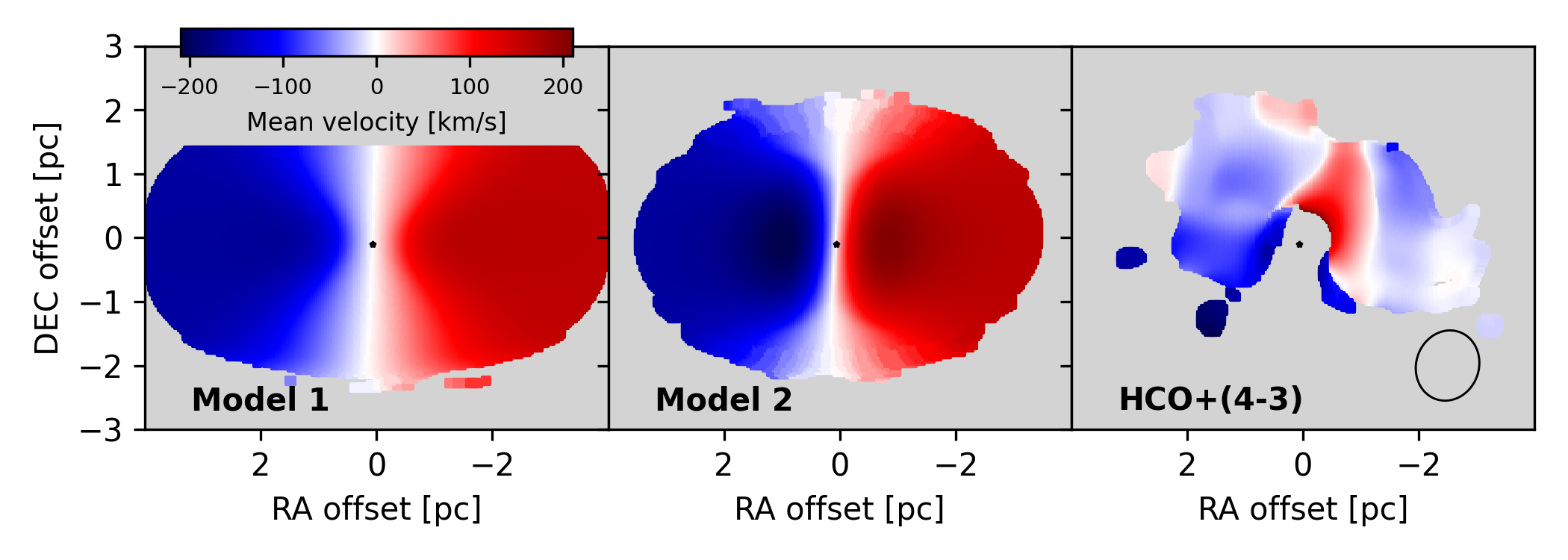}\vspace{-1.4cm}
   \includegraphics[width=1\hsize]{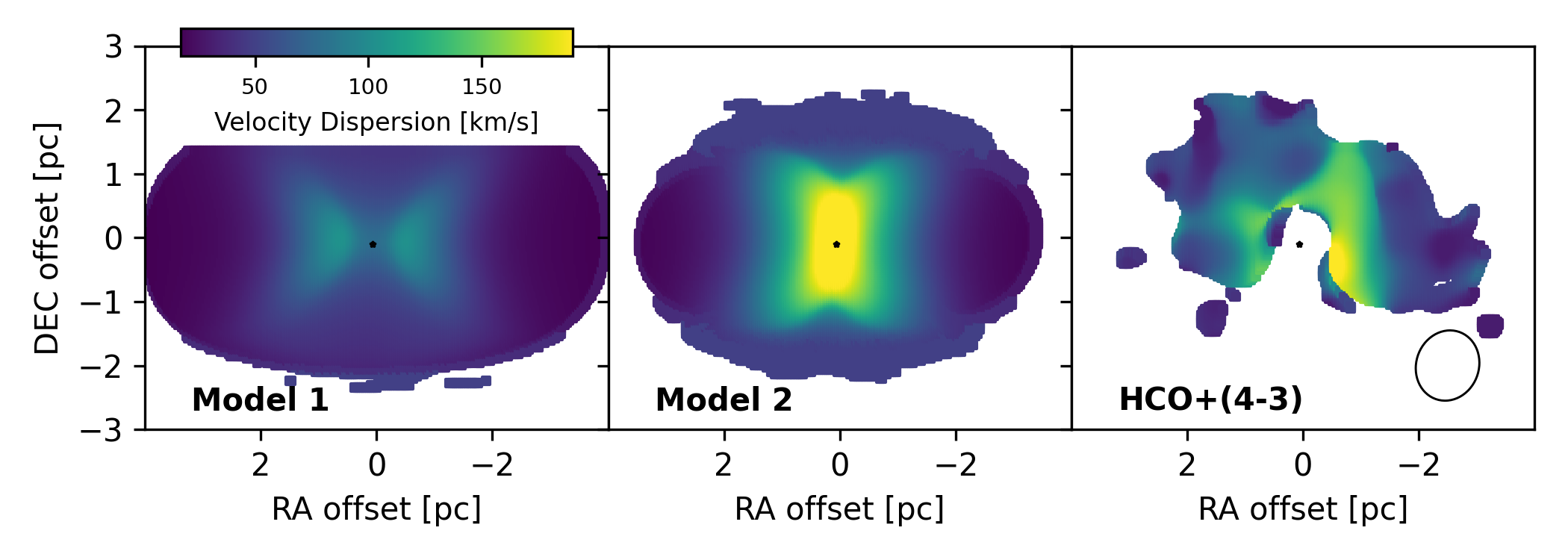}
    \caption[width=1\hsize]{Intensities at the mean velocities (top row), mean velocities (middle row) and  linewidths (bottom row) obtained from the Gaussian fit for Model 1 (left Col.), Model 2 (middle Col.) and the \rottrans{HCO$^+$}{4}{3} (right Col.). For all figures: the ellipse in the right panel represents the size of the beam for the observations, which we used to convolve the images of the models and the colour bar on top of the left panel applies to the three sub-images. The coordinate system is rotated by 24\degr{} (see details in text). 
    } \label{Fig:CompMod}
   \end{figure*}

Figure~\ref{Fig:CompMod} compares the integrated intensity, velocity field, and linewidths of the 3DBarolo simulations and our ALMA observations of \rottrans{HCO$^+$}{4}{3}. Note that the images have been rotated by 24\degr{} for comparison. It is difficult to compare the integrated intensity images because 3DBarolo does not simulate absorption; even so, the distribution of molecular gas is clearly more complex than our necessarily simplified 3DBarolo simulations. The simulated velocity field only broadly resembles that observed within the inner pc. Again, absorption challenges the comparison: only at the region immediately surrounding the central continuum source do we see rotation speeds comparable to the simulation. Possibly, the molecular accretion disc does not extend far beyond the region sampled by \water{} masers.

The linewidth simulations are more informative. In the simulations, the linewidths form a butterfly-shaped pattern within the central 2~pc. In Model 2, where the emissivity profile is more centrally concentrated, the linewidths are comparable to those observed in \rottrans{CO}{3}{2} and \rottrans{HCO$^+$}{4}{3}. We conclude that the observed high linewidths in the central 2~pc could be the result of beam smearing and plausibly trace unresolved rotation in the \water{} maser disc rather than turbulent motions.

\subsection{Asymmetric drift and sub-Keplerian rotation}\label{sec:asymmetricdrift}
The relatively high values of $\sigma/v$ seen in both the models and the observations suggest another explanation of the apparent observed sub-Keplerian rotation velocities, namely 'asymmetric drift'.  This is the phenomenon in axisymmetric systems, also seen in our Galaxy \citep{2008gady.book.....B}, where the  mean tangential velocity, $v_t$, is lower than the circular velocity, $v_c$, if the velocity dispersions are high and gradients in the density or dispersion exist.  The underlying cause is that in the presence of radial dispersions at any radius that we observe, we are also seeing clouds whose mean radius is smaller than the observed radius. From the conservation of angular momentum, these clouds will have tangential velocities lower than the circular velocities at this point.  This possibility was also noted by \cite{2022A&A...665A.102V}. The behaviour is captured in the axisymmetric Jeans Equation:
$$\frac{v_a}{v_c}= \frac{1}{2 v^2_c}\left (
\sigma^2_t-\sigma^2_r \left (1+
\frac{\partial \log \rho}{\partial\log r}+\frac{\partial\log \sigma^2_r}{\partial\log r}\right)\right ),$$
where $v_a\equiv <v_t>-v_c$ is the asymmetric drift; $\sigma_t,\ \sigma_r$ are the tangential and radial velocity dispersions, and $\rho$ is the cloud density; all quantities are potentially functions of $r$.  The observations suggest that, at least at some radii $\sigma/v_c\sim 1$ and varies radially.

In the central region, $1<r< 2$ pc, we can estimate roughly from the 
observed moment maps (Figs. \ref{Fig:RADECMeanVel} and \ref{Fig:RADECVelDisp}) that $\sigma\sim v_c\sim 100$ km/s,
$ \frac{\partial \log \sigma}{\partial\log r}<-2$, and we assume $ \frac{\partial \log \rho}{\partial\log r}<0$.
For isotropic turbulence ($\sigma_t\equiv\sigma_r\equiv\sigma$), this yields $v_a/v_c\sim 0.5$, so that the observed velocity would be about half the  actual Keplerian circular velocity.  In Sect. \ref{sec:lbG} we point out that for highly centrally concentrated gas distributions ($\frac{\partial \log \rho}{\partial\log r}<-2$) the beam smearing at the ALMA resolution causes an overestimation of $\sigma$ in the central region. However, in this case the larger value of the second to last term in the Jeans Equation compensates for the smaller value of the last term and leads to a similar estimate of $\frac{v_a}{v_c}$.

In the outer 2-4 pc region the situation is different.  $\sigma$ is approximately constant at $\sim 50$ km/s while $\partial \log \rho/\partial\log r\sim -1$, and $v_c\sim 140$ km/s, so the Jeans equation predicts $v_a/v_c\sim\sigma^2/2v_c^2\sim 0.04$, which is negligible.  The measured rotational velocities in these two regions are $\sim 90$ km/s and $<10$ km/s respectively, so the asymmetric drift can be considered a plausible explanation for the sub-Keplerian rotation in the inner region, but not at radii $r\frac{>}{\sim} $ 3 pc. The very low velocities at the latter projected positions cannot maintain the gas against the gravitational force of the central black hole. They may be transient, or at much larger radii but seen in projection closer to the nucleus, or supported by another mechanism, such as radiation pressure. 

\cite{ 2022A&A...665A.102V} present a numerical model of a dense cloud falling onto a pre-existing disc at the centre of NGC~1068, based on detailed assumptions on the physical processes in the parsec scale region. Their basic physical picture is quite similar to that presented here (discussed in Sects. \ref{sec:fil} and \ref{sec:kinmod} of this work) and fits the earlier, lower spatial resolution, ALMA data reasonably well. Here however, we have limited our analysis to quantities that can be derived fairly directly from the observed data, without additional assumptions about the dynamic processes.

\subsection {CO to HCO$^+$ line ratios}
\label{sec:lineratiodiscussion}

In Sect. \ref{sec:lineratios} we pointed out that the \rottrans{CO}{3}{2} to \rottrans{HCO$^+$}{4}{3} line ratios in the parsec-sized region around the nucleus are in the range 0.5--2 which is a factor of 10--20 lower than that seen in the CND, and also around the AGN when measured at much lower spatial resolution.  

\cite{2022A&A...665A.102V} presents molecular line images of the same transitions of other ALMA observations of NGC~1068 that show similar low ratios of the same lines and transitions to our, at a slightly lower spatial resolution. From their Table 4 we get an integrated intensity ratio of the CO and HCO$^+$ of 1.8 in lower-resolution observations of the torus region. \cite{2022A&A...665A.102V} did not comment explicitly on the ratios, nevertheless this value is consistent with the detailed physical/dynamical model presented in the same work within a factor of two or three (see also Sect. \ref{sec:agnspectra} below).  

\cite{GarciaBurillo2021} compiled line ratios for a large sample of AGNs measured at the AGN position. These are rather small, ranging from $\sim$2.8--5.7 for what they call the GATOS sample, and was interpreted as pointing at the presence of high-density molecular gas at small radii inside the central $\sim$3-10 pc. In comparison, another sample of AGNs, the NUGA sample \citep{Combes2019, Audibert2019}, shows higher values for the ratios of $\sim$8.5--38. An important difference between these two samples is that the latter have lower luminosity and Eddington ratios. The highest ratios around 30, were interpreted as an indication of the molecular torus having the lowest gas density. Nevertheless, outflows might explain the observations of the GATOS sample, having preferentially removed diffuse, lower-density molecular gas but leaving heavier and higher-density clumps behind. 

The difference by an order of magnitude in this line ratio was further interpreted as evidence of radial density stratification of the molecular tori \citep{GarciaBurillo2021}. Therefore, a drop in this ratio as one approaches the nucleus seems to be a more general trend in AGNs. 

In the diffuse interstellar medium in our Galaxy, the \rottrans{CO}{1}{0}/\rottrans{HCO$^+$}{1}{0} ratio is typically $\sim 100$ \citep{2023ApJ...943..172L}, much higher than the ratio for these lines  in the circumnuclear regions of AGNs (e.g. \cite{Viti2014}). 

This decrease in the CO/HCO$^+$ ratio near AGNs is usually attributed to higher densities \citep{GarciaBurillo2021}, because the critical density of the HCO$^+$ line ($\sim2\times10^6$ cm$^{-3}$) is much higher than that of CO ($\sim $8000 cm$^{-3}$). The critical density is that where collisional de-excitation of the upper state of a molecular transition competes with spontaneous radiative de-excitation; further density increases no longer cause increases in the molecular emission. However the calculations in \cite{Viti2014} (Table 12) indicate that the density in the more extended regions around the nucleus of NGC~1068 is already $\sim 10^6$ cm$^{-3}$, that is, at the HCO$^+$ critical density, so further density increases would not decrease the CO/HCO$^+$ ratio to the values we measure. In any case, the maximum line ratio shift between low and high-density environments would be the ratio of the critical densities of the two molecules, in this case $\sim 120$, so our observed minimum line ratio of $\sim 0.5$ is at the extreme edge of what can be explained by density effects, suggesting that a chemical shift in the abundance ratio of CO to HCO$^+$ also plays a role.  

\cite{2005AIPC..783..203K} attributed drops in the CO/HCN line ratio near AGNs to the intense X-ray environment near the nucleus. \cite{Izumi2013} and \cite{Martin2021} however claim that the low value of this ratio is more likely the result of high temperature induced chemical shifts near AGNs than X-ray or cosmic ray radiation. The same mechanisms may be responsible for the decrease of the CO/HCO$^+$ ratio from CND towards the AGN seen here. If future higher-resolution observations show that this ratio continues to drop closer to the nucleus, then mechanisms other than density increases must be involved.

\subsection{Near-nuclear motions and possible filamentary structures}\label{sec:fil}
\label{sec:noncircular}  In Sect. \ref{sec:agnspectra} we indicated the existence of a highly redshifted absorption feature seen as a dark blue region in Fig. \ref{Fig:pvmajco} at velocities of +350 to +500 km/s at the continuum peak position. The feature seems to be contiguous with emission to the west of the nucleus at velocities of -50 to +300 km/s.
The continuity in space and velocity suggest that this is one structure changing from emission to absorption as it passes in front of the AGN continuum. If the continuum source were in fact point-like this would indicate a non-circular, infalling cloud or filament, since material in circular motion would have the systemic velocity when seen in front of the source. However
\cite{2019ApJ...884L..28I} show that the brightest nuclear continuum emission is extended approximately 1.0 parsec, mostly along the major axis of our molecular emission. The redshifted absorption could in principle arise from circularly moving gas on the west side of the source, unresolved within our ALMA beam.

However, several arguments can be made against a simple model of a ring of molecular gas orbiting at a radius of $\sim 0.5$ parsec:
(1) The major-axis PV-diagram (See Fig. \ref{Fig:pvmajco}) is strongly asymmetric at velocities $|V-V_{sys}|>100$ km/s.
(2) When the ring crosses in front of the centre of the continuum source it should cause a deep absorption at systemic velocity.  In the data only a weak absorption is seen at this velocity.
(3) In a complete circular ring orbiting the continuum emitter, the highest velocity should be seen in emission, at the tangent point where the gas is moving directly away from the viewer, rather than absorption in front of the continuum source, where the velocity is diminished by the cosine of the angle to the line of sight.

These arguments suggest that the high-velocity redshifted feature seen in both emission and absorption represents an incomplete ring or filament primarily on the west side of the nucleus.  This material may also represent a filament or sheared cloud of gas on a highly elliptical or parabolic orbit.  In either case its motion would cause changes in appearance and velocity on timescales of $\sim r/v\sim 10^3$ years. We note that \cite{2009ApJ...691..749M} found a redshifted infalling filament or 'tongue' of molecular Hydrogen from the east on larger scales and lower velocities. Also, \cite{2019A&A...632A..61G} noted a large low-surface-brightness, low-velocity connection east of the nucleus to the CND.  

It is clear that the scenario of an infalling cloud does not connect with the interpretation of the outflow in HCN \citep{2019ApJ...884L..28I} nor with the inferred outflow at larger scales (CND scales, \cite{GarciaBurillo2014}) seen in several other molecular lines (CO(3–2), CO(6–5), HCN(4–3), and HCO$^{+}$(4–3)). Nevertheless these pictures are not mutually exclusive. The most intriguing puzzle is why we do not see the deep absorption at systemic velocity as in the HCN \citep{2019ApJ...884L..28I}. Nevertheless, the molecules CO and HCO$^{+}$ and the HCN, do not have a common chemical path, therefore they are not expected to behave in the same way. 

Whatever the kinematic origin of the molecular gas, the proximate appearance of absorption and emission in the nuclear spectra allow estimates of the physical conditions in the cloud. In Sect. \ref{sec:agnspectra} we estimated the peak continuum brightness temperature to be $T_{c}\sim 550$ K and the depth of the absorption feature to be $\Delta T_{a}\sim -30$ K while the off-nuclear line emission brightness fluxes  are $\sim 0.5-1.0$ mJy corresponding to brightness temperatures of $T_{em}\sim 30-60$ K. 
For gas with an excitation temperature of $T_{x}$ these quantities are related by $$\Delta T_a=(T_x-T_c)\tau\ ;\ T_{em}=T_{x}\tau,$$ 
$$where\,\, T_{x} = \frac{T_c}{1-\Delta T_a/T_{em}}\simeq 270-360\ {\rm K}.$$
The HCO$^+$ profile yields similar values. These are relatively high temperatures for molecular clouds but are similar to the dust temperatures found on similar scales in \cite{2022Natur.602..403G}. They may also indicate that the excitation temperature is influenced by the intense radiation field near the AGN. If we nonetheless make the assumption of LTE we can compute the column densities of the molecules.  The absorption profiles in Fig. \ref{Fig:abs} have  numerically integrated optical depths, $\int \tau dv$, of 20 km/s and 10 km/s for the \rottrans{CO}{3}{2} and \rottrans{HCO$^+$}{4}{3} transitions, respectively.  From LTE calculations at $T_x=300$ K, the total column densities in the two molecules are $N_{CO}\sim 5\times 10^{18} {\rm cm}^{-2}$ and $N_{HCO^+}\sim 3\times 10^{15} {\rm cm}^{-2}$.
For an abundance ratio $X_{HCO^+}\equiv N_{HCO^+}/N_{H_2}\simeq 3\times 10^{-9}$  similar to that found in diffuse molecular gas in our Galaxy \citep{2023ApJ...943..172L} the latter column density implies $N_{H_2}\sim 1\times 10^{24}$ cm$^{-2}$.
However, as discussed in Sect. \ref{sec:lineratiodiscussion} the abundances of HCO$^+$  may be enhanced in the high ionisation environment of the AGN, which would lower the estimated Hydrogen column density.
An independent lower limit to the column density of an orbiting dusty molecular cloud can be made from a 'dusty Eddington' calculation of the balance between radiation pressure and gravity. If all of the AGN's ultraviolet radiation is absorbed by the dust at the cloud's exposed face, the outwards radiation force per unit area is $L_{UV}/4\pi cr^2$, while the inward gravitational force per unit area of the overlying column of gas is $G M_{AGN} m_H N_H/r^2$. 
Taking the estimates of the mass and bolometric luminosity of NGC~1068 used in \cite{2020A&A...634A...1G} a cloud can only be in a stable or infalling orbit if $N_H\sim 1\times 10^{24}$ cm$^{-2}$ or greater, suggesting that the HCO$^+$ abundance in NGC~1068 is similar to the diffuse Galactic value.  In this case the high values of the HCO$^+$/CO line ratios discussed in \ref{sec:lineratios} indicate a deficiency of CO rather than a surplus of HCO$^+$ and the effect of the high ionisation environment near the AGN would be to destroy CO by conversion to C$^+$. 

\subsection{Kinematic modelling of the high-velocity feature}\label{sec:kinmod}
The absorption feature seen at velocities $v_{los}$ of up to 500 km/s must be an infalling cloud near the nucleus. This velocity should be compared to the estimate of the Keplerian circular velocity at 10 mas radius (0.7 pc) of 315 km/s in \cite{Gallimore2023} based on water maser kinematics.  In a Keplerian gravitational field this corresponds to a velocity of free fall from infinity of 445 km/s.  Thus measuring $v_{los}\sim 500$ km/s implies that the gas must be closer than 8 mas (0.5 pc) to the black hole. This is consistent with being within the ALMA beam at the nuclear position (FWHP $\sim 15$ mas) and the fact that it must be in front of the nuclear continuum source (radius $\sim 10$ mas from \cite{Gallimore2023}).  This radius estimate is similar to the size of the torus seen in the infrared \citep{2022Natur.602..403G} and the absorbing cloud is probably related to that structure. 

We have attempted to model the kinematics of the western high-velocity redshifted emission and absorption features as an extended filament falling in from infinity on a parabolic orbit.  For simplicity we have assumed that the pole of the orbit lies in the plane of the sky. In reality  the orbit of the filament must be somewhat inclined with respect to the that of the torus in order to avoid a collision between the two structures. With these assumptions there are two parameters that define an orbit: the angle of the symmetry axis of the parabola from the line of sight and the distance of closest approach to the black hole (the  {\it peritrepa}, where  {\it maeri trepa} is Greek for 'black hole'). The model orbits were chosen to reproduce the main features of the observed filament in the major-axis PV-diagram (Fig. \ref{Fig:pvmajco}):
A continuous feature approaching the nucleus from the west seen in emission at velocities of +80 to +300 km/s that turns into absorption at +300 to +500 km/s and is not seen as a blueshifted feature to the east of the nucleus.  We have ignored features within 80 km/s of systemic in the modelling because these are confused with features arising from the molecular disc at larger radii. We have only modelled the kinematics of the infalling material and the relative roles of absorption and emission as it passes in front of the central continuum source.  We have not included any detailed physics about the filament such changes in temperature, density or ionisation as it approaches the centre.

\begin{figure*}
\centering
    \includegraphics[height=10cm]{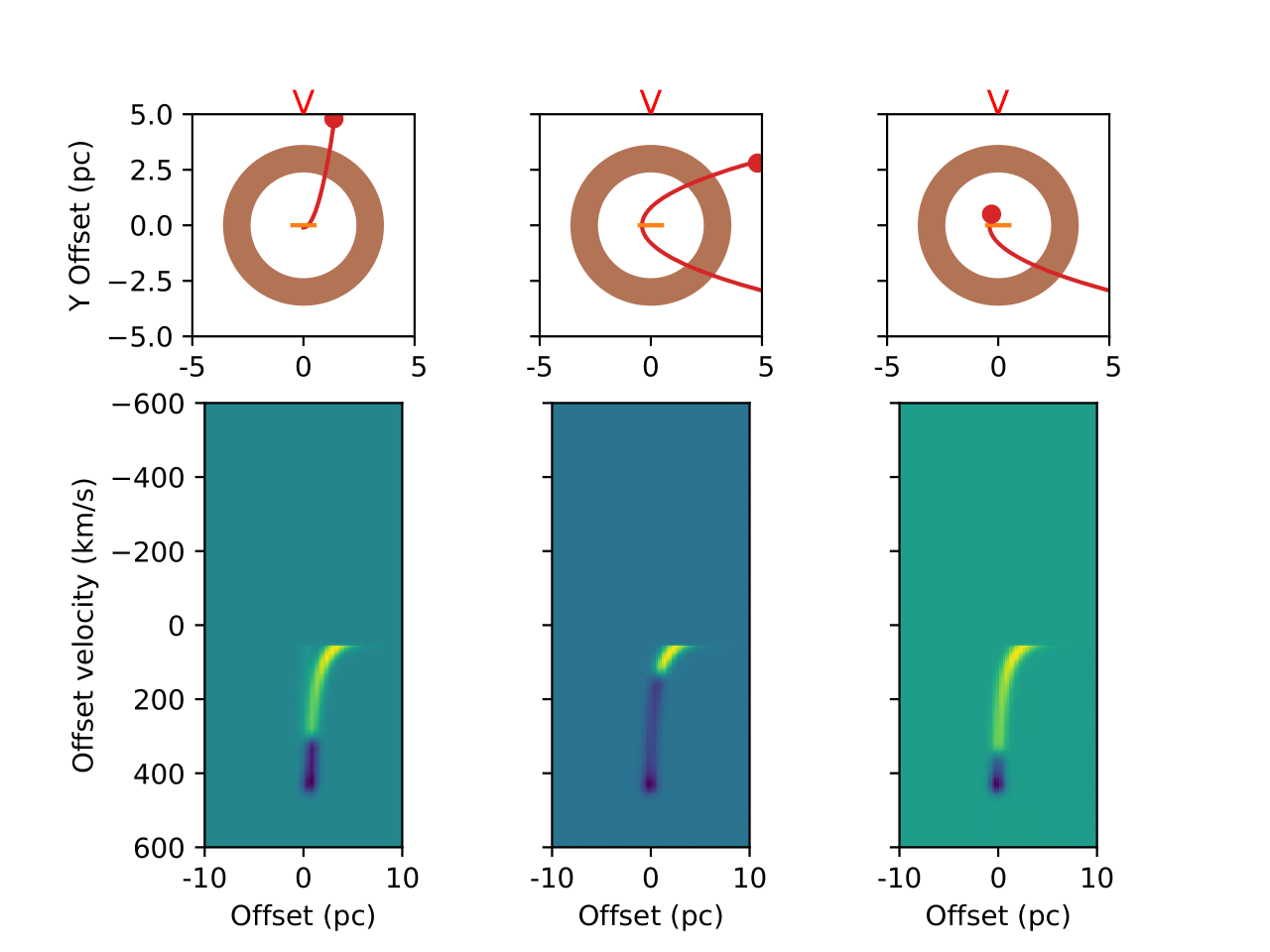} 
    \caption{The parabolic trajectories of a filament of clouds approaching the  peritrepa in three simulations. The top drawings illustrate the orbits in the plane of the filament, while the bottom series illustrate synthetic PV-diagrams derived from the system of filament plus continuum source that can be compared with the actual observations in Fig. \ref{Fig:pvmajco}. In each case the observer views the system from the top (red 'V' mark). The orange bar indicates the continuum source; clouds in front of this from the observer's point of view are seen in absorption, otherwise in emission. The red dot indicates where the orbit enters the field of view.  The brown circle at 3 pc radius indicates the typical radius of the molecular disc described in this paper.  To avoid a collision between the disc, the orbital plane of the filament must be somewhat inclined with respect to that of the disc. The 3-dimensional rendering of the first model in Fig. \ref{Fig:3Dorb} illustrates the relative orientations of the disc, filament, continuum source and observer for this case.   The three models are:{\bf Left:}  a trajectory seen along the parabolic axis.  Distance of closest approach is 0.1 pc (see text) and the orbit is truncated at this point. {\bf Centre:} the parabolic axis is perpendicular to the line-of-sight. {\bf Right:} Same as {\bf Centre} but the filament only begins just before peritrepa. Velocities within $\pm 80$ km/s of systemic have been suppressed.  The synthetic PV-distributions have been convolved with a Gaussian of 0.5 pc in the spatial direction to simulate the ALMA resolution.}
    \label{Fig:orbits}
\end{figure*}

\begin{figure*}
\centering
    \includegraphics[height=6cm]{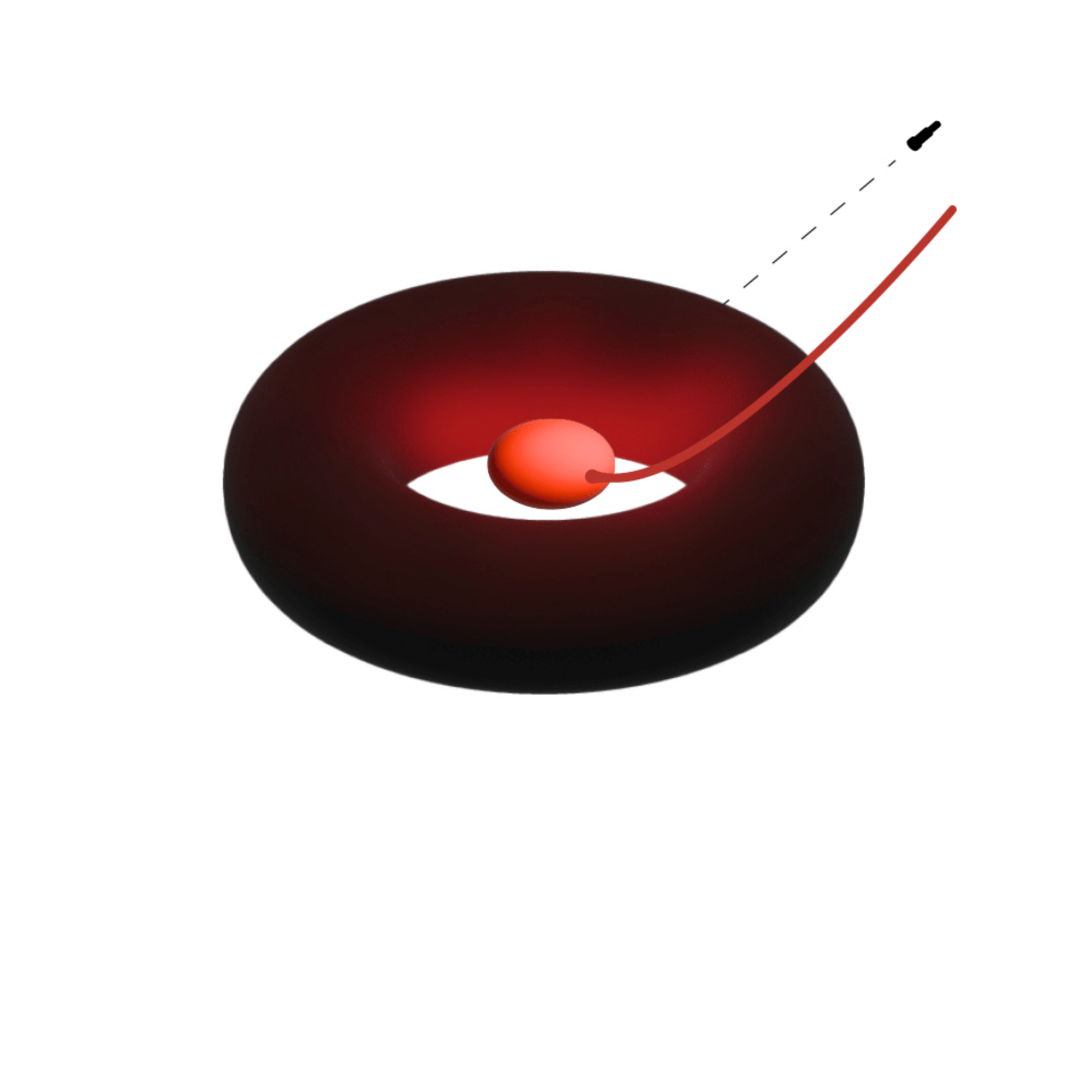}
    \caption{A 3-dimensional rendering showing the spatial relationships of the components for the first case in Fig. 10 (Left), the observer's line of sight from the back of the illustration, and the inclination of the disc plane and the filament's orbital plane.  Illustration: W. Westerik.} 
    \label{Fig:3Dorb}
\end{figure*}

Figure \ref{Fig:orbits} shows the trajectories in the plane of the orbits and simulated PV-diagrams for a string of clouds along three model orbits; Fig. \ref{Fig:3Dorb} shows a 3-dimensional rendering of the first model to aid visualization. Figure \ref{Fig:orbits} {\bf Left} shows a model where the axis of the orbit is along the line-of-sight and peritrepa=0.1 pc; {\bf Centre} shows a model where it is perpendicular to the line of sight and peritrepa=0.5 pc. {\bf Right} is similar to {\bf Centre} but with the initial part of the orbit truncated. Comments on these three models:
{\bf Left} matches the emission/absorption structure of Fig. \ref{Fig:pvmajco}.  It was necessary to terminate the orbit at closest approach in order to eliminate the appearance of a similar blueshifted eastern feature. This termination could be explained by destruction of the clouds near the black hole by radiation near the central source. To reproduce the maximum observed velocity, $\sim 500$ km/s, the distance of closest approach must be quite small, 0.1 pc because most of the velocity at this point is projected perpendicular to the line of sight.  This model requires the continuum source to be extended so that some of the filamentary clouds are seen in absorption before they pass around the back of the black hole. In fact this source has a radius $\sim 0.7$ pc \citep{Gallimore2023}), larger than the orbital peritrepa distance.
{\bf Centre} shows an orbit approaching from the west, passing 0.4 pc from the centre while moving directly away from the observer and leaving again to the west.  The peritrepa distance is larger because the total velocity at this point is along the line of sight, and the velocity is redshifted along the entire orbit so there is no problem eliminating an eastern blueshifted component.  There is excess absorption at relatively low velocities relative to the observed PV-diagram because the clouds are seen in front of the continuum source for a relatively long path while at larger radii and moving across the line of sight. In this model the actual size of the continuum source is not important.
{\bf Right} The excess absorption in the {\bf centre} model can be suppressed by clipping the piece of the orbit  in front of the continuum source. This change improves the kinematic match to the observed PV-diagram but of course leaves unanswered the problem of why we don't see the initial, infalling, segment of the orbit.

The larger radius of $\sim 0.4$ pc is similar to that of the main infrared dust structure.  A radius of $\sim 0.1$ pc would put the molecular clouds in the filament at or within the 'sublimation radius' where the UV-radiation from the central engine destroys the dust. This radius was estimated by \cite{2020A&A...634A...1G} to be in the range 0.08-0.27 pc, depending on the assumed total UV-luminosity of NGC~1068.  In all three cases the peritrepa radius is  within the estimated radius of the continuum emitting source
\citep{1996ApJ...464..198G}, but we do not here further explore the consequences of this proximity.

\section{Conclusions}
We present high spatial and spectral resolution ALMA observations of the \rottrans{HCO$^+$}{4}{3} and  \rottrans{CO}{3}{2} lines in the near-nuclear region of the prototype Seyfert 2 galaxy NGC~1068 and its underlying continuum. We present the results as moment maps, PV-diagrams, and spectra at the position of the nuclear continuum source. We made a simple kinematic model with help of the software 3DBarolo and we analysed the possibility of explaining the non-Keplerian motions with asymmetric drift. We discuss the absorption feature seen in both lines on top of the AGN and present a simple model for a cloud in an elliptical orbit. We also notice the very low line intensity ratio of the CO to the HCO$^{+}$.
We arrive at the following conclusions:
\begin{itemize}
    \item The line emitting regions traced by CO and $HCO^{+}$ (see Fig. \ref{Fig:overplot}) present significant asymmetry in morphologies, mean velocities, surface brightness, and linewidths, when comparing the east and west sides of the disc, suggesting a highly disturbed kinematic system. Such traces can be a consequence of possible past events (e.g. mergers) or ongoing processes (accretion or jet interactions).
    
    \item In the inner regions (radius $\leq$ 2 pc) the sub-Keplerian rotation can be explained by asymmetric drift associated with the high velocity dispersions. This is crucial for understanding the stability and longevity of the molecular structures around the AGN. Nevertheless, the very low velocity components at larger radii (r$\geq$3 pc) cannot be explained this way, radiation pressure may also be important to support the gas. In particular, in the east side the \rottrans{CO}{3}{2} emission seems to extend out to $\sim13$ pc, from radii of $\sim4$ pc; this is most probably a filament connecting the torus to the CND. 

    \item The gas very close to the AGN shows very broad linewidths which may indicate high turbulent velocities there. A simple kinematic model indicates that the broad lines seen in the inner $\leq$2 pc can also be accounted for beam smearing of the Keplerian disc associated to the H$_2$O masers with a high central density. 
    
    \item The molecular gas near the central continuum source or S1, observed in both emission at approximately 300 km/s and redshifted absorption at around 400 km/s, suggests the presence of dynamic and possibly transient structures, such as a partial ring, filament, or sheared gas clouds on elliptical or parabolic orbits. These features exhibit high-excitation temperatures around 300 K and significant column densities (e.g. \(N_{\text{CO}} \approx 5 \times 10^{18} \text{ cm}^{-2}\), \(N_{\text{HCO}^+} \approx 3 \times 10^{15} \text{ cm}^{-2}\)). The high velocity of the absorption component places it at or within the infrared dusty structure at 0.5 pc from the central engine, while some of the kinematic models place it much closer, $\sim 0.1$ pc, near the sublimation radius of the dust, suggesting that they are intimately connected to the central engine, possibly marking the innermost regions of AGN feeding channels where extreme physical processes are at play.

    \item The observed CO/HCO$^+$ line ratios in NGC 1068 are notably lower than typically seen in other AGNs, and there is a clear trend towards lower ratios with increasing spatial resolution. This pattern likely reflects the proximity of the gas to the AGN, where intense X-ray and cosmic ray emissions from the central engine drive the chemical conversion of CO into CI or CII. This transformation reduces CO abundances in regions illuminated directly by the AGN, suggesting a direct influence of the AGN's radiation on the surrounding molecular gas chemistry.
\end{itemize}

 The high resolution ALMA data presented here reveal highly asymmetric and dynamic molecular gas structures, including inflows and turbulent motions, that must play crucial roles in both fuelling the AGN and in the feedback processes that regulate the host galaxy's evolution.

\begin{acknowledgements}
    This paper makes use of the following ALMA data: ADS/JAO.ALMA 2019.1.01540.S. ALMA is a partnership of ESO (representing its member states), NSF (USA) and NINS (Japan), together with NRC (Canada), NSTC and ASIAA (Taiwan), and KASI (Republic of Korea), in cooperation with the Republic of Chile. The Joint ALMA Observatory is operated by ESO, AUI/NRAO and NAOJ.
    SGB acknowledges support from the Spanish grant \emph{PID2022-138560NB-I00}, funded by MCIN/AEI/10.13039/501100011033/FEDER, EU. 
    IM acknowledges financial support from the \emph{Severo Ochoa grant CEX2021-001131-S} funded by MCIN/AEI/10.13039/501100011033, and the \emph{Spanish MCIU grant PID2022-140871NB-C21}.
    CRA acknowledges support from the Agencia Estatal de Investigaci\'on of the Ministerio de Ciencia, Innovaci\'on y Universidades (MCIU/AEI) under the grant ``Tracking active galactic nuclei feedback from parsec to kiloparsec scales'', with reference PID2022$-$141105NB$-$I00 and the European Regional Development Fund (ERDF).
    The PI acknowledges assistance from Allegro, the European ALMA Regional Center node in the Netherlands. We thank Ed Formalont for valuable advice on the imaging process of the data. The authors thank Willem Westerik for providing graphic support for Fig. 11.

\end{acknowledgements}

%
%
\bibliographystyle{aa}
\bibliography{N1068}

\end{document}